# Multidimensional HLLE Riemann Solver; Application to Euler and Magnetohydrodynamic Flows

By
Dinshaw S. Balsara (dbalsara@nd.edu)
Physics Department, University of Notre Dame

**Abstract**

In this work we present a general strategy for constructing multidimensional HLLE Riemann solvers, with particular attention paid to detailing the two-dimensional HLLE Riemann solver. This is accomplished by introducing a constant resolved state between the states being considered, which introduces sufficient dissipation for systems of conservation laws. Closed form expressions for the resolved fluxes are also provided to facilitate numerical implementation. The Riemann solver is proved to be positively conservative for the density variable; the positivity of the pressure variable has been demonstrated for Euler flows when the divergence in the fluid velocities is suitably restricted so as to prevent the formation of cavitation in the flow.

We also focus on the construction of multidimensionally upwinded electric fields for divergence-free magnetohydrodynamical (MHD) flows. A robust and efficient second order accurate numerical scheme for two and three dimensional Euler and MHD flows is presented. The scheme is built on the current multidimensional Riemann solver and has been implemented in the author's RIEMANN code. The number of zones updated per second by this scheme on a modern processor is shown to be cost competitive with schemes that are based on a one-dimensional Riemann solver. However, the present scheme permits larger timesteps.

Accuracy analysis for multidimensional Euler and MHD problems shows that the scheme meets its design accuracy. Several stringent test problems involving Euler and MHD flows are also presented and the scheme is shown to perform robustly on all of them.

**1) Introduction**

Riemann solvers have long been recognized as being an important building block for robust and accurate schemes for conservation laws. Consequently, much attention has been lavished in the computational fluid dynamics community on the design of efficient Riemann solvers. Exact Riemann solvers for Euler flow have been formulated by Godunov [29] and van Leer [51]. While van Leer [51] had originally presented an efficient Newton iteration procedure for evaluating the exact Riemann problem for Euler flow, several authors have tried to build more efficient approximate Riemann solvers. The premise underlying this enterprise is that much of the information provided by the Riemann solver is indeed never used in the construction of a numerical flux. Thus there is the two-shock Riemann solver of Colella [15], the two-rarefaction fan Riemann solver of Osher and Solomon [39], the linearized Riemann solver by Roe [41], the HLLE Riemann solver (Harten, Lax & van Leer [31], Einfeldt [21]) and the HLLC Riemann solvers (Einfeldt *et al.* [22], Toro, Spruce and Speares [50], Batten *et al.* [10]). All of the above-mentioned Riemann solvers resolve the discontinuity at a zone boundary into a one-dimensional foliation of waves.

Despite this spate of rather good one-dimensional Riemann solvers, some practitioners have always believed that the one-dimensional Riemann solvers lose much of their efficacy in multidimensional problems. Indeed the motivation for thinking so stems from the fact that these Riemann solvers cannot account for flow features that might be propagating transverse to the zone boundary. The one-dimensional Riemann solvers are, therefore, biased to pick up flow variations that are orthogonal to the zone faces of a computational mesh. It is believed that the directional bias that is built into one-dimensional Riemann solvers causes a reduction in the permissible Courant number in multidimensional flow. Consequently, some practitioners have attempted to use the one-dimensional Riemann solvers in very intricate combinations in order to achieve multidimensional upwinding (Colella [16], Saltzman [48], LeVeque [35]). This form of multidimensional upwinding did indeed enable the design of schemes that operate with an increased Courant number, even though it sometimes came at the expense of solving a rather large number of one-dimensional Riemann problems.

Other researchers tried to build more complex models for multidimensional wave propagation (Roe [42], Rumsey, van Leer & Roe [45]). Early successes emerged with the work of Abgrall [1], [2] and have been followed up by the work of other practitioners (Fey [24], [25], Gilquin, Laurens & Rosier [28], Brio, Zakharian & Webb [13]). These authors attempted to obtain a multidimensional analogue of Roe's linearized Riemann solver. Thus a mean state was chosen at each edge of the computational mesh and the linearized, two-dimensional Euler equations were evolved in space and time to obtain a multidimensional solution of the Riemann problem. While elegant, this procedure only works for the Euler equations. It cannot be applied to any other hyperbolic system without a substantial amount of reformulation. The HLLE Riemann solver, on the other hand, works transparently for any conservation law. This prompted an early attempt by Wendroff [52] to formulate a multidimensional HLLE Riemann solver. Wendroff's formulation introduced nine constant states, which made his scheme unwieldy. To keep it tractable, he had to artificially expand the signal speeds to handle supersonic situations, which further increased dissipation. By contrast, the multidimensional HLLE Riemann solver that is presented here is much simpler and naturally accommodates all the supersonic cases. The flux calculation is also much simpler in this work, yielding closed form expressions for the fluxes that are easily implemented on a computer. Moreover, the positivity of our multidimensional HLLE Riemann solver is easily demonstrated, whereas such a demonstration eluded Wendroff. Consequently, one of the goals of this paper is to present our multidimensional HLLE Riemann solver and provide sufficient amount of detail to facilitate numerical implementation. We also wish to demonstrate the performance of this Riemann solver when it is applied to the computation of multidimensional Euler flows. We, therefore, present details of a spatially and temporally second order accurate scheme for Euler and magnetohydrodynamic (MHD) flows that uses our multidimensional HLLE Riemann solver as a building block.

In recent years it has become interesting to apply techniques drawn from higher order Godunov schemes for hydrodynamics to other hyperbolic systems of conservation laws. These systems are usually larger and more complicated and an evaluation of their eigenstructure carries a greater computational complexity. Thus methods for constructing Riemann solvers that do not rely on evaluating the eigenvectors are favored. The MHD system provides a case in point. Ever since the analysis of the MHD eigensystem by Roe & Balsara [43] it has become possible to design robust, efficient, one-dimensional, linearized Riemann solvers for numerical MHD (Cargo and Gallice [14], Balsara [3]). An HLLC Riemann solver, capable of capturing mesh-aligned contact discontinuities, was presented by Gurski [30]. Miyoshi and Kusano [37] drew on Gurski's work to design an HLLD Riemann solver for MHD. In addition to contact discontinuities, the HLLD Riemann solver was also capable of capturing mesh-aligned Alfven waves. It is, therefore, one of the goals of this paper to present the performance of the multidimensional HLLE Riemann solver on multidimensional MHD problems.

A multidimensional Riemann solver has a utility in numerical MHD that goes beyond the construction of upwinded fluxes. The magnetic fields in the MHD system satisfy the property that they remain divergence-free for all time. Brackbill & Barnes [11] have shown that violating the divergence-free aspect of the magnetic field leads to unphysical plasma transport orthogonal to the magnetic field. One possible resolution is to formulate constrained transport schemes (Brecht *et al.* [12], DeVore [20], Evans & Hawley [23]) which collocate magnetic fields at zone centers and use edge-centered electric fields for their divergence-free update. Another solution might be to modify the MHD equations (Powell [40]) at the expense of introducing source terms in the momentum and energy update equations, thus relinquishing momentum and energy conservation. Dedner *et al.* [19] have formulated another kind of modification of the MHD system where the divergence that builds up in the magnetic field is propagated away at a predetermined signal speed. Soon after the advent of higher order Godunov schemes for MHD, Dai & Woodward [18], Ryu *et al.* [47] and Balsara & Spicer [9] formulated higher order Godunov methods that kept the magnetic field divergence-free. The essential idea in Balsara & Spicer [9] was to rely on the dualism between certain components of the upwinded flux vector that is evaluated at zone faces and the electric fields that are sought at the zone edges. By using the facially upwinded fluxes to obtain the edge-centered electric field components, the previous authors introduced a modicum of upwinding into the evaluation of the electric field. However, Balsara & Spicer [9] were acutely aware of the need for multidimensional upwinding and provide a whole section in their paper on that issue. Subsequent efforts have drawn on the same dualism between the electric fields and flux components. However, recent work has tried to increase the amount of dissipation from one-dimensional Riemann solvers to obtain more stable multidimensional upwinding (Londrillo and DelZanna [36], Gardiner & Stone [27]). There has also been work on using a genuinely divergence-free reconstruction and collocating the one-dimensional Riemann solvers at zone edges (Balsara [5], [6], Balsara *et al.* [7]) so that a more accurate representation of the electric field can be obtained. It is, however, difficult to know what this multidimensional upwinding at the zone edges ought to be without having a multidimensional Riemann solver for MHD in hand. The multidimensional Riemann solver presented here is general enough to be applied to any hyperbolic system of conservation laws, including MHD. Since our multidimensional HLLE Riemann solver is applied at zone edges and yields two sets of upwinded fluxes, our further goal in this paper is to use it to obtain properly upwinded electric fields at zone edges for MHD calculations.

Multidimensional Riemann solvers have sometimes been perceived as being too complicated or difficult to implement. Part of this perception stems from the fact that most users of traditional higher order Godunov codes are accustomed to a dimension by dimension approach for building the update terms. This simplifies the scheme by requiring the fluxes to be evaluated at facial boundaries. The present Riemann solver, like most multidimensional Riemann solvers, is implemented at zone edges, which does require a slight paradigm shift in the implementer's thinking. The fluxes in the present Riemann solver are, however, very easy to build, requiring no more evaluations than those that would have been made for a one-dimensional HLLE Riemann solver. The method presented here easily extends to any hyperbolic conservation law. The approach in this paper is shown to work on a large number of stringent test problems in two and three dimensions. In return for the slightly greater complexity of implementation, the resulting scheme can operate with larger CFL numbers. The multidimensional Riemann solver also provides an unambiguous evaluation of the multidimensionally upwinded electric field in divergence-free formulations for MHD.

The plan of this paper is as follows. Section 2 presents the multidimensional HLLE Riemann solver. Section 3 examines multidimensional upwinding as it applies to computing edge-centered electric fields in MHD. Section 4 briefly describes the second order accurate predictor-corrector scheme that uses the multidimensional Riemann solver described here. Section 5 presents an accuracy analysis for Euler and MHD flows, showing that the schemes meet their design accuracy. Section 6 presents some stringent multidimensional test problems drawn from Euler flow. Section 7 does the same for MHD flow. Section 8 presents conclusions. In an Appendix we show that our Riemann solver always produces positive densities for Euler and MHD flows. In that same Appendix we also demonstrate the pressure positivity of our Riemann solver for Euler flows when the divergence of the velocity lies in certain ranges so as to exclude multidimensional cavitations.

**2) Multidimensional HLLE Riemann Solver**

We divide this section into three sub-sections. Sub-section 2.1 gives the derivation of the multidimensional HLLE Riemann solver. Sub-section 2.2 provides an analogous multidimensional LLF Riemann solver. Sub-section 2.3 gives details on how the multidimensional fluxes are to be assembled at zone faces and considers the restrictions placed on the timestep.

**2.1) Derivation of the Multidimensional HLLE Riemann Solver**

Consider an *N*-component system of conservation laws

$$\partial_t \mathbf{U} + \partial_x \mathbf{F} + \partial_y \mathbf{G} = 0 .\qquad(1)$$

Here $\mathbf{U}$ is the vector of conserved variables and $\mathbf{F}$ and $\mathbf{G}$ are the flux vectors in the x and y-directions. Say that we want to formulate a multidimensional HLLE solver on Cartesian meshes, or on any logically rectangular mesh. Fig. 1 shows a schematic diagram of such a situation where the four zones that come together at an edge are shown by the four quadrants of the coordinate system. The edge itself is located at the origin *O*. The initial conditions for this multidimensional Riemann problem consist of four constant states, $\mathbf{U}_{RU}$, $\mathbf{U}_{LU}$, $\mathbf{U}_{LD}$ and $\mathbf{U}_{RD}$, in the first, second, third and fourth quadrants respectively, as shown in Fig. 1. A mnemonic strategy for remembering the subscripts *RU, LU, LD* and *RD* is given in the figure caption of Fig. 1. We assume that we can identify the largest right and left-going wave speeds emerging from that edge and denote them by $S_R$ and $S_L$. These could be obtained, say for instance, by considering two x-directional HLLE Riemann solvers, one located immediately above the x-axis and another immediately below the x-axis. Thus the speeds $S_R$ and $S_L$ represent the maximal right and left-going speeds obtained from both those Riemann solvers. We can similarly identify the largest upward and downward-going wave speeds emerging from the same edge and denote them by $S_U$ and $S_D$. Einfeldt [21] and Batten *et al.* [10] provide prescriptions for obtaining these one-dimensional extremal speeds, and the same can be used here in multi-dimensions. Thus let $\lambda_x^1(\mathbf{U}_{RU})$ and $\lambda_x^N(\mathbf{U}_{RU})$ denote the smallest and largest x-directional wave speeds respectively in the state $\mathbf{U}_{RU}$ with corresponding definitions for the other states. Let $\bar{\lambda}_x^1(\mathbf{U}_{LU}, \mathbf{U}_{RU})$ and $\bar{\lambda}_x^N(\mathbf{U}_{LU}, \mathbf{U}_{RU})$ be the smallest and largest x-directional wave speeds from a linearized Riemann solver that is applied between the states $\mathbf{U}_{LU}$ and $\mathbf{U}_{RU}$ with similar definitions for the other pairs of states. Make similar definitions for the y-direction. The extremal speeds $S_R$, $S_L$, $S_U$ and $S_D$ are then given by

$$\begin{aligned}
S_R &= \max\left( \lambda_x^N(\mathbf{U}_{RU}),\ \lambda_x^N(\mathbf{U}_{RD}),\ \bar{\lambda}_x^N(\mathbf{U}_{LU}, \mathbf{U}_{RU}),\ \bar{\lambda}_x^N(\mathbf{U}_{LD}, \mathbf{U}_{RD}) \right) \\
S_L &= \min\left( \lambda_x^1(\mathbf{U}_{LU}),\ \lambda_x^1(\mathbf{U}_{LD}),\ \bar{\lambda}_x^1(\mathbf{U}_{LU}, \mathbf{U}_{RU}),\ \bar{\lambda}_x^1(\mathbf{U}_{LD}, \mathbf{U}_{RD}) \right) \\
S_U &= \max\left( \lambda_y^N(\mathbf{U}_{RU}),\ \lambda_y^N(\mathbf{U}_{LU}),\ \bar{\lambda}_y^N(\mathbf{U}_{RD}, \mathbf{U}_{RU}),\ \bar{\lambda}_y^N(\mathbf{U}_{LD}, \mathbf{U}_{LU}) \right) \\
S_D &= \min\left( \lambda_y^1(\mathbf{U}_{RD}),\ \lambda_y^1(\mathbf{U}_{LD}),\ \bar{\lambda}_y^1(\mathbf{U}_{RD}, \mathbf{U}_{RU}),\ \bar{\lambda}_y^1(\mathbf{U}_{LD}, \mathbf{U}_{LU}) \right)
\end{aligned} \qquad(2)$$

Dissipation is produced in the one-dimensional HLLE Riemann solver by assuming a constant state that lies between the left and right states. As long as the extremal speeds are based on a physically sound choice, the presence of this constant state introduces the requisite amount of dissipation. In multiple dimensions, this domain will likely be a circle or ellipse. However, in the interest of a simple formulation, we assume that the cell-break problem is started at $t = 0$ and that by a time $t = T$ the constant state fills the rectangle shown in Fig. 1. The left panel of Fig. 2 shows the simple wave model that we adopt for the propagation of waves in two spatial dimensions and one temporal direction. Since this wave model circumscribes the actual waves that propagate out from the initial discontinuity, it will provide adequate amount of dissipation in multi-dimensions. Later we show that it can even account for all the supersonic cases in multiple dimensions.

As in the one-dimensional HLLE Riemann solver, we wish to identify the fluxes at the boundaries of the space-time domain being considered. Associated with the constant state $\mathbf{U}_{RU}$ in the first quadrant, we can evaluate the x and y-fluxes $\mathbf{F}_{RU}$ and $\mathbf{G}_{RU}$ which prevail at the line segments *MA* and *MC* respectively, see Fig. 1. The constant state $\mathbf{U}_{LU}$ in the second quadrant yields the fluxes $\mathbf{F}_{LU}$ and $\mathbf{G}_{LU}$ at the line segments *NB* and *NC* respectively. One can draw on the constant state $\mathbf{U}_{LD}$ in the third quadrant to evaluate the fluxes $\mathbf{F}_{LD}$ and $\mathbf{G}_{LD}$ which correspond to the line segments *RB* and *RD* respectively. Likewise, the constant state $\mathbf{U}_{RD}$ in the fourth quadrant yields the fluxes $\mathbf{F}_{RD}$ and $\mathbf{G}_{RD}$ at the line segments *QA* and *QD* respectively.

For the one-dimensional HLLE Riemann solver, the constant resolved state, $\mathbf{U}^*$, is obtained by carrying out a two-dimensional integration of the conservation law in space-time. The derivation of that resolved state is most easily obtained if one begins by considering the subsonic case. For that reason, we start our derivation by assuming the situation where $S_L < 0 < S_R$ and $S_D < 0 < S_U$. (We will show how this assumption is relaxed before the end of this Section.) For our present spatially two-dimensional problem, the constant resolved state, $U^*$, can be obtained by integrating the conservation law, eqn. (1), over a three dimensional rectangular prism in space-time. The base of this rectangular prism at $t = 0$ is given by the rectangle *QMNR* in Fig. 1. The set of vertices that make up this rectangular prism in space-time is given in the $(x, y, t)$ coordinate system by $\{(S_R T, S_U T, 0), (S_L T, S_U T, 0), (S_L T, S_D T, 0), (S_R T, S_D T, 0), (S_R T, S_U T, T), (S_L T, S_U T, T), (S_L T, S_D T, T), (S_R T, S_D T, T)\}$. Integrating eqn. (1) over this rectangular prism gives (after an obvious cancellation of a factor of $T^2$)

$$\mathbf{U}^*(S_R - S_L)(S_U - S_D) - \mathbf{U}_{RU} S_R S_U + \mathbf{U}_{RD} S_R S_D + \mathbf{U}_{LU} S_L S_U - \mathbf{U}_{LD} S_L S_D$$
$$+ (\mathbf{F}_{RU} - \mathbf{F}_{LU})S_U - (\mathbf{F}_{RD} - \mathbf{F}_{LD})S_D + (\mathbf{G}_{RU} - \mathbf{G}_{RD})S_R - (\mathbf{G}_{LU} - \mathbf{G}_{LD})S_L = 0 \quad (3)$$

The above equation can now be written as

$$\mathbf{U}^* = \frac{\mathbf{U}_{RU} S_R S_U + \mathbf{U}_{LD} S_L S_D - \mathbf{U}_{RD} S_R S_D - \mathbf{U}_{LU} S_L S_U}{(S_R - S_L)(S_U - S_D)}$$
$$- \frac{(\mathbf{F}_{RU} - \mathbf{F}_{LU})S_U - (\mathbf{F}_{RD} - \mathbf{F}_{LD})S_D + (\mathbf{G}_{RU} - \mathbf{G}_{RD})S_R - (\mathbf{G}_{LU} - \mathbf{G}_{LD})S_L}{(S_R - S_L)(S_U - S_D)} \quad (4)$$

The previous equation yields the resolved state for the multidimensional Riemann problem.

It is interesting to observe several aspects of the resolved state given in eqn. (4). First, notice that when the variation is confined to be in the x-direction we have

$$\mathbf{U}_{RU} = \mathbf{U}_{RD} \text{ and } \mathbf{U}_{LU} = \mathbf{U}_{LD} \Rightarrow \mathbf{F}_{RU} = \mathbf{F}_{RD}, \mathbf{F}_{LU} = \mathbf{F}_{LD}, \mathbf{G}_{RU} = \mathbf{G}_{RD} \text{ and } \mathbf{G}_{LU} = \mathbf{G}_{LD} \quad (5)$$

In that limit we obtain

$$\mathbf{U}^* = \frac{S_R \mathbf{U}_{RU} - S_L \mathbf{U}_{LU} - (\mathbf{F}_{RU} - \mathbf{F}_{LU})}{S_R - S_L}. \tag{6}$$

Eqn. (6) is just the familiar formula for the resolved state of the one-dimensional HLLE Riemann solver in the subsonic case. We see, therefore, that in the subsonic case our multidimensional Riemann solver produces the expected resolved state when all the variations are one-dimensional. A similar reduction occurs when all the variations are confined to the y-direction. Notice though that in all other situations the resolved state always picks up multidimensional variations that are not contained in the one-dimensional resolved state. This is as one would expect for a multidimensional Riemann solver.

We now focus on obtaining the resolved x and y-fluxes, $\mathbf{F}^*$ and $\mathbf{G}^*$ respectively, for our multidimensional Riemann solver. Again, we first restrict attention to the subsonic case but we will relax that assumption later on. Our derivation will become easier if we make the following simplifying definitions, each of which has a self-evident meaning within the context of the one-dimensional HLLE Riemann solver. Thus we define

$$\mathbf{F}_U^{\text{HLLE}} \equiv \left[\frac{S_R}{S_R - S_L}\right] \mathbf{F}_{LU} - \left[\frac{S_L}{S_R - S_L}\right] \mathbf{F}_{RU} + \left[\frac{S_R S_L}{S_R - S_L}\right] (\mathbf{U}_{RU} - \mathbf{U}_{LU}), \tag{7}$$

$$\mathbf{F}_D^{\text{HLLE}} \equiv \left[\frac{S_R}{S_R - S_L}\right] \mathbf{F}_{LD} - \left[\frac{S_L}{S_R - S_L}\right] \mathbf{F}_{RD} + \left[\frac{S_R S_L}{S_R - S_L}\right] (\mathbf{U}_{RD} - \mathbf{U}_{LD}), \tag{8}$$

$$\mathbf{G}_R^{\text{HLLE}} \equiv \left[\frac{S_U}{S_U - S_D}\right] \mathbf{G}_{RD} - \left[\frac{S_D}{S_U - S_D}\right] \mathbf{G}_{RU} + \left[\frac{S_U S_D}{S_U - S_D}\right] (\mathbf{U}_{RU} - \mathbf{U}_{RD}), \tag{9}$$

and

$$\mathbf{G}_L^{\text{HLLE}} \equiv \left[\frac{S_U}{S_U - S_D}\right] \mathbf{G}_{LD} - \left[\frac{S_D}{S_U - S_D}\right] \mathbf{G}_{LU} + \left[\frac{S_U S_D}{S_U - S_D}\right] (\mathbf{U}_{LU} - \mathbf{U}_{LD}). \tag{10}$$

We will try to express the resolved x and y-fluxes in terms of the above four familiar definitions because they also make it easy to identify those parts of the fluxes that carry the dissipation terms and those parts of the fluxes that carry the genuinely multidimensional contributions.

To obtain $\mathbf{F}^*$, the resolved x-flux from the multidimensional Riemann solver, we have again to integrate the conservation law from eqn. (1) over a three-dimensional rectangular prism in space-time. The base of this rectangular prism at $t = 0$ is given by the rectangle *QMCD* in Fig. 1. The set of vertices that make up this rectangular prism in space-time is given in the $(x, y, t)$ coordinate system by $\{( S_R T, S_U T, 0), ( 0, S_U T, 0), ( 0, S_D T, 0), ( S_R T, S_D T, 0), ( S_R T, S_U T, T), ( 0, S_U T, T), ( 0, S_D T, T), ( S_R T, S_D T, T)\}$. To simplify the derivation, we first evaluate the area integral on the $x = 0$ face. Notice that the multidimensional wave shown in Fig. 2 will not reach all points on that face. As a result, we only have two x-directional Riemann problems operating on those parts of the $x = 0$ face that are not overtaken by the multidimensional wave. The $x = 0$ face is shown separately in the right panel of Fig. 2 along with the fluxes that propagate through parts of that face. The area integral of the x-flux on the $x = 0$ face is, therefore, given by

$$\frac{1}{2} S_U T^2 \mathbf{F}_U^{HLLE} - \frac{1}{2} S_D T^2 \mathbf{F}_D^{HLLE} + \frac{1}{2} (S_U - S_D) T^2 \mathbf{F}^* . \tag{11}$$

Integrating the conservation law, eqn. (1), over the rectangular prism that we have identified in this paragraph yields ( after an obvious cancellation of a factor of $T^2$ )

$$\mathbf{U}^* (S_U - S_D) S_R - \mathbf{U}_{RU} S_R S_U + \mathbf{U}_{RD} S_R S_D + \mathbf{F}_{RU} S_U - \mathbf{F}_{RD} S_D$$
$$- \frac{1}{2} \mathbf{F}_U^{HLLE} S_U + \frac{1}{2} \mathbf{F}_D^{HLLE} S_D - \frac{1}{2} \mathbf{F}^* (S_U - S_D) + (\mathbf{G}_{RU} - \mathbf{G}_{RD}) S_R = 0 \tag{12}$$

After a certain amount of algebraic simplification eqn. (12) yields

$$\mathbf{F}^* = \frac{\mathbf{F}_{LU} S_R S_U + \mathbf{F}_{RD} S_L S_D - \mathbf{F}_{LD} S_R S_D - \mathbf{F}_{RU} S_L S_U}{(S_R - S_L)(S_U - S_D)}$$
$$- 2 \left[ \frac{S_R S_L}{(S_R - S_L)(S_U - S_D)} \right] (\mathbf{G}_{RU} - \mathbf{G}_{LU} + \mathbf{G}_{LD} - \mathbf{G}_{RD}) \tag{13}$$
$$+ \left[ \frac{S_R S_L}{(S_R - S_L)(S_U - S_D)} \right] \left[ S_U (\mathbf{U}_{RU} - \mathbf{U}_{LU}) - S_D (\mathbf{U}_{RD} - \mathbf{U}_{LD}) \right]$$

Eqn. (13) is useful because it allows us to pick out the contributions from the x and y-fluxes as well as the dissipation term. We see that the first term in eqn. (13) is a convex combination of x-fluxes. This is the non-dissipative part of the x-flux. The second term contains the contribution from the y-fluxes to the resolved x-directional flux. Any genuinely multidimensional Riemann solver should include such contributions from the fluxes in the transverse direction. Observe that the contribution from the y-fluxes to eqn. (13) depends on the relative magnitudes of $S_R$ and $S_L$. Thus the amount of transverse flux contributed to the resolved x-flux depends on the direction and speed with which signals propagate in the x-direction. Notice though that for smooth flow, the contribution from the y-fluxes is small. The last term in eqn. (13) contains the dissipation terms.

It is useful to observe that the dissipation terms in eqn. (13) only provide dissipation in the x-direction; there are no dissipation terms in the y-direction. At strong shocks, and especially at strong oblique shocks, there are no dissipation terms in eqn. (13) corresponding to the y-direction. At such shocks the contribution from the second term in eqn. (13), i.e. the terms with the y-fluxes, can become significant. There is no further y-directional dissipation term corresponding to the y-fluxes at strong shocks, as a result, the only strategy for stabilizing the x-flux at strong shocks is to have a shock-detector and truncate the second term in the vicinity of strong shocks. If the underlying numerical method has a shock detector (Colella & Woodward [17], Balsara *et al.* [7]) it pays to suppress the contribution of the transverse terms in the vicinity of shocks. Thus we introduce a flow-dependent parameter $\beta$ that smoothly goes to zero in the vicinity of a shock. At all other locations in the computation we set $\beta = 1$. The final form for the resolved flux is then written as

$$\mathbf{F}^* = \frac{\mathbf{F}_{LU}\,S_R S_U + \mathbf{F}_{RD}\,S_L S_D - \mathbf{F}_{LD}\,S_R S_D - \mathbf{F}_{RU}\,S_L S_U}{(S_R - S_L)(S_U - S_D)}$$
$$- 2\beta \left[\frac{S_R S_L}{(S_R - S_L)(S_U - S_D)}\right](\mathbf{G}_{RU} - \mathbf{G}_{LU} + \mathbf{G}_{LD} - \mathbf{G}_{RD}) \qquad (14)$$
$$+ \left[\frac{S_R S_L}{(S_R - S_L)(S_U - S_D)}\right]\left[S_U (\mathbf{U}_{RU} - \mathbf{U}_{LU}) - S_D (\mathbf{U}_{RD} - \mathbf{U}_{LD})\right]$$

By comparing eqns. (13) and (14), we see that the latter equation restricts the contribution of the transverse fluxes at strong shocks. It is interesting to point out that in their analysis of multidimensional schemes for scalar advection Roe and Sidilkover [44] also found that the transverse fluxes have to be restricted in certain situations. This completes our description of the method for obtaining the resolved flux $\mathbf{F}^*$ from the multidimensional HLLE Riemann solver.

A particularly compact and interesting form for the resolved flux $\mathbf{F}^*$ is given by

$$\mathbf{F}^* = \left[\frac{S_U}{S_U - S_D}\right] \mathbf{F}_U^{HLLE} - \left[\frac{S_D}{S_U - S_D}\right] \mathbf{F}_D^{HLLE}$$
$$- 2\beta \left[\frac{S_R S_L}{(S_R - S_L)(S_U - S_D)}\right] (\mathbf{G}_{RU} - \mathbf{G}_{LU} + \mathbf{G}_{LD} - \mathbf{G}_{RD})$$
(15)

This is also the form that is most useful for computations. Notice that when the variation is confined to be in the x-direction, eqn. (15) reduces appropriately. Consequently, for x-directional variations we have $\mathbf{F}^* = \mathbf{F}_U^{HLLE} = \mathbf{F}_D^{HLLE}$. For genuinely multidimensional problems notice two important features of eqn. (15). First, observe that the first two terms of eqn. (15) represent a linear weighting of the regular HLLE fluxes $\mathbf{F}_U^{HLLE}$ and $\mathbf{F}_D^{HLLE}$. The weighting is simply proportional to the relative areas of the rectangles *BAMN* and *BAQR* from Fig. 1, which lends itself to a very simple geometrical interpretation. Second, we see that the third term of the resolved flux picks up extra contributions from the y-directional fluxes.

To obtain $\mathbf{G}^*$, the resolved y-flux from the multidimensional Riemann solver, we have to integrate the conservation law from eqn. (1) over a three-dimensional rectangular prism in space-time. The base of this rectangular prism at $t = 0$ is given by the rectangle *AMNB* in Fig. 1. We do not provide all the steps in the derivation of $\mathbf{G}^*$ because they are very similar to our previous derivation. We simply quote the final form of the resolved flux $\mathbf{G}^*$ in a compact form as

$$\mathbf{G}^* = \left[\frac{S_R}{S_R - S_L}\right] \mathbf{G}_R^{HLLE} - \left[\frac{S_L}{S_R - S_L}\right] \mathbf{G}_L^{HLLE}$$
$$- 2\beta \left[\frac{S_U S_D}{(S_R - S_L)(S_U - S_D)}\right] (\mathbf{F}_{RU} - \mathbf{F}_{LU} + \mathbf{F}_{LD} - \mathbf{F}_{RD})$$
(16)

Just like eqn. (15), eqn. (16) lends itself to an elegant physical interpretation.

Up to this point we have focused on the transonic case. There is a very simple way in which the flux formula is generalized to handle the supersonic case in the one-dimensional HLLE Riemann solver. The structure of eqns. (7) to (10) suggests that the same might work here. Thus we assert that eqns. (15) and (16) easily extend to all possible supersonic cases if the signal speeds are reset as

$$S_L \to \min(S_L, 0) \;,\; S_R \to \max(S_R, 0) \;,\; S_D \to \min(S_D, 0) \;,\; S_U \to \max(S_U, 0). \qquad (17)$$

We consider three interesting examples below:
1) Say we originally start with a situation where $S_L > 0$ and $S_D > 0$. It corresponds to a physical problem where all the waves in the Riemann problem are propagating supersonically into the first quadrant. The use of eqn. (17) in eqns. (15) and (16) and also eqns. (7) to (10) then yields $\mathbf{F}^* = \mathbf{F}_{LU}$ and $\mathbf{G}^* = \mathbf{G}_{RD}$, i.e. the Riemann solver indeed picks out the correct upwinded fluxes that would be contributing to the first quadrant. Notice, quite interestingly, that this is not obtained by an arithmetic averaging of the fluxes at the four faces that come together at the edge. Later in this paper we will see that this example also yields insights into the process of obtaining edge-centered electric fields in schemes for divergence-free MHD.

2) Let us consider another example where we originally have $S_L < 0 < S_R$ and $S_D > 0$. The use of eqn. (17) in eqn. (15) then shows us that $\mathbf{F}^*$ does not depend any more on $\mathbf{F}_D^{HLLE}$, which is as one would expect from the upwinding. Notice though that $\mathbf{F}^*$ still depends on $\mathbf{F}_U^{HLLE}$. Furthermore, $\mathbf{F}^*$ continues to depend on the y-fluxes, though their contribution diminishes as $S_U$ increases.

3) In this example, let us consider a situation where we originally start with $S_L > 0$ and $S_D < 0 < S_U$. Now $\mathbf{F}^*$ depends exclusively on $\mathbf{F}_{LU}$ and $\mathbf{F}_{LD}$, i.e. it picks up the requisite upwinding in the x-direction. Notice, however, that the y-fluxes do not contribute to $\mathbf{F}^*$.

The Appendix demonstrates that the Riemann solver designed here keeps the density positive for Euler and MHD flows. We also demonstrate that when the velocities are restricted so as to preclude multidimensional cavitations in the flow, our Riemann solver keeps the pressure positive.

**2.2) Multidimensional LLF Riemann Solver**

It is also possible to obtain an LLF (or Rusanov [46]) variant of the fluxes in eqns. (13) and (16) by setting

$$S \equiv \max\left(|S_R|, |S_L|, |S_U|, |S_D|\right) \ ; \ S_R \to S \ ; \ S_L \to -S \ ; \ S_U \to S \ ; \ S_D \to -S \quad (18)$$

to get

$$\begin{aligned}\mathbf{F}^* = &\frac{1}{4}\left(\mathbf{F}_{RU} + \mathbf{F}_{LU} + \mathbf{F}_{RD} + \mathbf{F}_{LD}\right) + \frac{1}{2}\beta\left(\mathbf{G}_{RU} - \mathbf{G}_{LU} + \mathbf{G}_{LD} - \mathbf{G}_{RD}\right) \\ &- \frac{S}{4}\left(\mathbf{U}_{RU} - \mathbf{U}_{LU} + \mathbf{U}_{RD} - \mathbf{U}_{LD}\right)\end{aligned} \quad (19)$$

and

$$\begin{aligned}\mathbf{G}^* = &\frac{1}{4}\left(\mathbf{G}_{RU} + \mathbf{G}_{RD} + \mathbf{G}_{LU} + \mathbf{G}_{LD}\right) + \frac{1}{2}\beta\left(\mathbf{F}_{RU} - \mathbf{F}_{LU} + \mathbf{F}_{LD} - \mathbf{F}_{RD}\right) \\ &- \frac{S}{4}\left(\mathbf{U}_{RU} - \mathbf{U}_{RD} + \mathbf{U}_{LU} - \mathbf{U}_{LD}\right)\end{aligned} \quad (20)$$

The first terms in eqns. (19) and (20) contain the contribution from the dissipation-free x and y-fluxes respectively at the edge being considered. The second terms in eqns. (19) and (20) contain the contribution from the fluxes in the transverse direction. The last terms in eqns. (19) and (20) contain the dissipation terms.

## 2.3) Assembling the Multidimensional Fluxes at Zone Faces and Timestep Considerations

The previous sub-sections have provided the derivation of the multidimensional HLLE Riemann solver without specifying how one should assemble the final flux at each zone face of a two-dimensional zone. As shown in Fey [24], Brio, Zakharian and Webb [13] and Kurganov, Noelle and Petrova [34] the multidimensional flux has to be assembled at a zone face by considering the contributions coming from a one-dimensional Riemann solver evaluated at the center of the zone face and the multidimensional Riemann solvers evaluated at the corners of that face. Consider the zone in Fig. 3 and say that its sides have a length of $\Delta x$ and $\Delta y$ in the x and y-directions. The zone is denoted by indices $(i, j)$ with appropriate half-integer notational extensions to denote zone faces and corners. Fig. 3 shows the evolution of Riemann problems at all the faces and all the corners of the two-dimensional zone for a time $\Delta t$. In other words, at each zone face we also solve a one-dimensional Riemann problem in addition to solving the multidimensional Riemann problem at each corner. As time $\Delta t$ increases, the multidimensional Riemann problems at each of the corners make an increasingly larger contribution to the facially and temporally averaged fluxes at the zone faces. This is especially true in the subsonic cases shown in Fig. 3. These multidimensional contributions have a beneficial and stabilizing effect on the one-dimensional flux, because they represent the contribution from the cross-terms that arise when making a Taylor expansion of the original partial differential equation. With the help of Fig. 3 we can arrive at a space-time averaged version of the flux $\bar{\mathbf{F}}_{i+1/2,j}$ at the $(i+1/2, j)$ face. As shown in Fig. 3, $\mathbf{F}^*_{i+1/2,j}$ is the resolved x-flux coming from the one-dimensional Riemann solver at the $(i+1/2, j)$ face, $\mathbf{F}^*_{i+1/2, j+1/2}$ is the resolved x-flux from the multidimensional Riemann solver at the corner $(i+1/2, j+1/2)$ and $\mathbf{F}^*_{i+1/2, j-1/2}$ is the resolved x-flux from the multidimensional Riemann solver at the corner $(i+1/2, j-1/2)$. The final expression for the multidimensionally corrected, space-time averaged x-flux is given by

$$\bar{\mathbf{F}}_{i+1/2,j} = \mathbf{F}^*_{i+1/2,j} \left[ 1 - \left( S_{U, i+1/2, j-1/2} - S_{D, i+1/2, j+1/2} \right) \frac{\Delta t}{2\Delta y} \right]$$
$$- \mathbf{F}^*_{i+1/2, j+1/2} \, S_{D, i+1/2, j+1/2} \frac{\Delta t}{2\Delta y} + \mathbf{F}^*_{i+1/2, j-1/2} \, S_{U, i+1/2, j-1/2} \frac{\Delta t}{2\Delta y} \quad . \tag{21}$$

Eqn. (21) is so designed that it extends seamlessly to the supersonic limits when eqn. (17) is applied to the wave speeds.

We see from Fig. 3 that even in the extreme limit where the waves emanating from the multidimensional Riemann solvers at $(i+1/2, j+1/2)$ and $(i+1/2, j-1/2)$ touch each other at the x-face, the flux $\bar{\mathbf{F}}_{i+1/2,j}$ gets a contribution of at least $\mathbf{F}^*_{i+1/2,j}/2$ from the one-dimensional Riemann solver. This is inevitable considering that eqn. (21) is a space-time average evaluated over the x-face. The condition that the waves emanating from the corners of any face in Fig. 3 should not intersect each other is explicitly given by

$$\Delta t \leq \min\left( \frac{\Delta x}{S_{R,i-1/2,j+1/2} - S_{L,i+1/2,j+1/2}}, \frac{\Delta x}{S_{R,i-1/2,j-1/2} - S_{L,i+1/2,j-1/2}}, \frac{\Delta y}{S_{U,i+1/2,j-1/2} - S_{D,i+1/2,j+1/2}}, \frac{\Delta y}{S_{U,i-1/2,j-1/2} - S_{D,i-1/2,j+1/2}} \right). \tag{22}$$

In practice we might make the less restrictive requirement that $\bar{\mathbf{F}}_{i+1/2,j}$ should be a convex combination of the fluxes $\mathbf{F}^*_{i+1/2,j}$, $\mathbf{F}^*_{i+1/2,j+1/2}$ and $\mathbf{F}^*_{i+1/2,j-1/2}$ in eqn. (21). This allows us to double the timestep constraint in eqn. (22), yielding a maximum CFL number of unity. Thus the final constraint on the timestep for two-dimensional flow is given by

$$\Delta t \leq \min\left( \frac{2\Delta x}{S_{R,i-1/2,j+1/2} - S_{L,i+1/2,j+1/2}}, \frac{2\Delta x}{S_{R,i-1/2,j-1/2} - S_{L,i+1/2,j-1/2}}, \frac{2\Delta y}{S_{U,i+1/2,j-1/2} - S_{D,i+1/2,j+1/2}}, \frac{2\Delta y}{S_{U,i-1/2,j-1/2} - S_{D,i-1/2,j+1/2}} \right). \tag{23}$$

In practice, the timestep is evaluated using the conventional zone-centered approach. Eqn. (23) only serves to illustrate that a larger CFL number might be possible. This completes our description of timestep constraints in two dimensions.

There is, however, a deficiency in eqn. (21) that becomes apparent to those who are familiar with the old ENO schemes from the 1980s, see Harten *et al.* [32]. Notice that eqn. (21) changes form as the waves at the edge change direction and speed. Thus the weights ascribed to the fluxes $\mathbf{F}^*_{i+1/2,j}$, $\mathbf{F}^*_{i+1/2,j+1/2}$ and $\mathbf{F}^*_{i+1/2,j-1/2}$ in eqn. (21) keep changing. This is tantamount to having a rapidly changing stencil. As with the old ENO schemes, this results in a loss of accuracy in certain circumstances. For that reason, we prefer to incorporate the multidimensional Riemann solver using a Simpson rule which fixes the relative weights of the fluxes and yields

$$\bar{\mathbf{F}}_{i+1/2,j} = \frac{1}{6}\mathbf{F}^*_{i+1/2,j+1/2} + \frac{4}{6}\mathbf{F}^*_{i+1/2,j} + \frac{1}{6}\mathbf{F}^*_{i+1/2,j-1/2}. \tag{24}$$

Please note that eqn. (24) may relinquish some of the timestep advantages of eqn. (21), but in practice it permits timesteps that are quite large without ever degrading the order of accuracy. Eqn. (24) is formally third order accurate if the fluxes $\mathbf{F}^*_{i+1/2,j+1/2}$ and $\mathbf{F}^*_{i+1/2,j-1/2}$ are evaluated at the upper and lower x-face using the multidimensional Riemann solver and if the flux $\mathbf{F}^*_{i+1/2,j}$ is evaluated at the center of the x-face. Notice however that during the evaluation of $\mathbf{F}^*_{i+1/2,j+1/2}$ and $\mathbf{F}^*_{i+1/2,j-1/2}$ we do evaluate the x-directional HLL fluxes at the x-face. I.e. we are referring to the two HLLE fluxes at the x-face that are evaluated using just the variables and their moments in zones $(i,j)$ and $(i+1,j)$. If one is willing to accept second order accuracy then those two x-directional HLL fluxes can be averaged to the center of the x-face to yield a second order accurate approximation for $\mathbf{F}^*_{i+1/2,j}$. This is the economical choice that we made for the applications that we present later. Eqn. (24) is also trivially extended to three dimensions, especially when one only desires a second order scheme. Thus in three dimensions an easy way of evaluating an x-flux at the upper x-face of the zone $(i,j,k)$ while using the multidimensional Riemann solver at the edges of that face consists of extending eqn. (24) as

$$\overline{\mathbf{F}}_{i+1/2,j,k} = \frac{1}{6}\mathbf{F}^*_{i+1/2,j+1/2,k} + \frac{1}{6}\mathbf{F}^*_{i+1/2,j-1/2,k} + \frac{1}{6}\mathbf{F}^*_{i+1/2,j,k+1/2} + \frac{1}{6}\mathbf{F}^*_{i+1/2,j,k-1/2} + \frac{2}{6}\mathbf{F}^*_{i+1/2,j,k}. \quad (25)$$

Eqn. (25) and its analogues in the other two directions were used to evaluate the fluxes in all the three dimensional calculations shown here. As with eqn. (24), if we want a second order accurate scheme then $\mathbf{F}^*_{i+1/2,j,k}$ in eqn. (25) does not need to be evaluated but can be obtained via an averaging process.

**3) Multidimensional Upwinding of Edge-centered Electric Fields in MHD**

In Sub-section 3.1 we obtain explicit expressions for the electric field for MHD using the multidimensional Riemann solver from the previous section. In Sub-Section 3.2) we put the present work in context by comparing it to expressions for the electric field obtained from prior research.

**3.1) Electric Field Expressions from the Multidimensional HLLE Riemann Solver**

The three-dimensional MHD system can be written in flux conservation form, $\partial_t \mathbf{U} + \partial_x \mathbf{F} + \partial_y \mathbf{G} + \partial_z \mathbf{H} = 0$, as

$$\frac{\partial}{\partial t}\begin{pmatrix} \rho \\ \rho\, v_x \\ \rho\, v_y \\ \rho\, v_z \\ \varepsilon \\ B_x \\ B_y \\ B_z \end{pmatrix} + \frac{\partial}{\partial x}\begin{pmatrix} \rho\, v_x \\ \rho\, v_x^2 + P + \mathbf{B}^2/8\pi - B_x^2/4\pi \\ \rho\, v_x v_y - B_x B_y/4\pi \\ \rho\, v_x v_z - B_x B_z/4\pi \\ \left(\varepsilon + P + \mathbf{B}^2/8\pi\right) v_x - B_x(\mathbf{v}\cdot\mathbf{B})/4\pi \\ 0 \\ \left(v_x B_y - v_y B_x\right) \\ -\left(v_z B_x - v_x B_z\right) \end{pmatrix}$$

$$+ \frac{\partial}{\partial y}\begin{pmatrix} \rho\, v_y \\ \rho\, v_x v_y - B_x B_y/4\pi \\ \rho\, v_y^2 + P + \mathbf{B}^2/8\pi - B_y^2/4\pi \\ \rho\, v_y v_z - B_y B_z/4\pi \\ \left(\varepsilon + P + \mathbf{B}^2/8\pi\right) v_y - B_y(\mathbf{v}\cdot\mathbf{B})/4\pi \\ -\left(v_x B_y - v_y B_x\right) \\ 0 \\ \left(v_y B_z - v_z B_y\right) \end{pmatrix} + \frac{\partial}{\partial z}\begin{pmatrix} \rho\, v_z \\ \rho\, v_x v_z - B_x B_z/4\pi \\ \rho\, v_y v_z - B_y B_z/4\pi \\ \rho\, v_z^2 + P + \mathbf{B}^2/8\pi - B_z^2/4\pi \\ \left(\varepsilon + P + \mathbf{B}^2/8\pi\right) v_z - B_z(\mathbf{v}\cdot\mathbf{B})/4\pi \\ \left(v_z B_x - v_x B_z\right) \\ -\left(v_y B_z - v_z B_y\right) \\ 0 \end{pmatrix} = 0$$

(26)

where $\rho$ is the density; $v_x$, $v_y$ and $v_z$ are the velocity components; $B_x$, $B_y$ and $B_z$ are the magnetic field components; $\varepsilon = \rho\, v^2/2 + P/(\gamma - 1) + \mathbf{B}^2/8\pi$ is the total energy and $\gamma$ is the ratio of specific heats. Note though that the divergence-free update equation for the magnetic field is still given by

$$\frac{\partial \mathbf{B}}{\partial t} + c\, \nabla \times \mathbf{E} = 0 \; ; \; \mathbf{E} \equiv -\frac{1}{c}\, \mathbf{v} \times \mathbf{B} \tag{27}$$

where $\mathbf{E}$ is the electric field vector. The speed of light "c" cancels out in eqn. (27). Consequently, for the sake of simplicity, we do not carry it in the ensuing equations. Balsara and Spicer [9] realized that there is a dualism between the fluxes that are produced by a higher order Godunov scheme and the electric fields that were needed in eqn. (27). We see that the flux components of eqn. (26) obey the following symmetries:

$$E_x = -G_8 = H_7 \; ; \; E_y = F_8 = -H_6 \; ; \; E_z = -F_7 = G_6 \; . \tag{28}$$

The last three components of the **F**, **G** and **H** fluxes could also be reinterpreted as electric fields in the dual approach. The electric fields are needed at the edge centers as shown in Fig. 4 and are to be used to update the face-centered magnetic fields. For example, on a Cartesian mesh with zone sizes $\Delta x$, $\Delta y$ and $\Delta z$ a one-step, second order accurate discretization of the x-component of the magnetic field in eqn. (27) yields

$$\bar{B}^{n+1}_{x;\,i+1/2,j,k} = \bar{B}^{n}_{x;\,i+1/2,j,k} - \frac{\Delta t}{\Delta y \Delta z}\left(\Delta z \bar{E}^{n+1/2}_{z;\,i+1/2,j+1/2,k} - \Delta z \bar{E}^{n+1/2}_{z;\,i+1/2,j-1/2,k} + \Delta y \bar{E}^{n+1/2}_{y;\,i+1/2,j,k-1/2} - \Delta y \bar{E}^{n+1/2}_{y;\,i+1/2,j,k+1/2}\right).$$
(29)

Here $\bar{B}^{n}_{x;\,i+1/2,j,k}$ and $\bar{B}^{n+1}_{x;\,i+1/2,j,k}$ are the facially-averaged magnetic fields at times $t^n$ and $t^{n+1} = t^n + \Delta t$ and the time-centered electric field components $\bar{E}^{n+1/2}_{y;\,i+1/2,j,k-1/2}$ and $\bar{E}^{n+1/2}_{z;\,i+1/2,j-1/2,k}$ are collocated as shown in Fig. 4. Details on implementing CT schemes that are based on higher order Godunov methodology are provided in [5] and [9]. Using these magnetic and electric fields in eqn. (29) and its analogues in the other two directions then yields a divergence-free, i.e. constrained transport, update strategy. In this section we focus on obtaining the upwinded forms of the z-component of the electric field at the z-edges of the zone shown in Fig. 4. Eqn. (28) shows us that we will, therefore, have to focus on $F_7$, the seventh component of the x-flux, and $G_6$, the sixth component of the y-flux.

Let us, therefore, explicitly write out eqn. (13) as it applies to the seventh component of the x-flux, i.e. $F_7^*$. We see immediately that the seventh component of the y-flux in eqn. (26) is zero. However, the multidimensional upwinding from the Riemann solver that was designed in the previous Section plays an important role in deciding how the electric fields from $F_7$ in eqn. (26) are to be combined. It also provides the structure of the dissipation terms. To obtain a good appreciation of the dissipation terms, realize from Fig. 1 that when a divergence-free reconstruction is used for the magnetic field (Balsara [4], [5]) $B_x$ is continuous between the states $\mathbf{U}_{LU}$ and $\mathbf{U}_{RU}$ while $B_y$ undergoes a jump between those two states. Fig. 1 also shows us that $B_x$ is continuous but $B_y$ undergoes a jump between the states $\mathbf{U}_{LD}$ and $\mathbf{U}_{RD}$. Thus the jumps in $B_y$ produce the dissipation terms in $F_7^*$. Using eqn. (13) we therefore get

$$F_7^* = -\frac{E_{z,LU}\,S_R S_U + E_{z,RD}\,S_L S_D - E_{z,LD}\,S_R S_D - E_{z,RU}\,S_L S_U}{(S_R - S_L)(S_U - S_D)}$$
$$+ \left[\frac{S_R S_L}{(S_R - S_L)(S_U - S_D)}\right]\left[S_U\left(B_{y,RU} - B_{y,LU}\right) - S_D\left(B_{y,RD} - B_{y,LD}\right)\right]$$
(30)

Similarly, realize that $B_y$ is continuous but $B_x$ undergoes a jump between the states $\mathbf{U}_{RD}$ and $\mathbf{U}_{RU}$. Likewise, $B_y$ is continuous but $B_x$ undergoes a jump between the states $\mathbf{U}_{LD}$ and $\mathbf{U}_{LU}$. Thus the jumps in $B_x$ produce the dissipation terms in $G_6^*$. Using eqn. (16) gives us

$$G_6^* = \frac{E_{z,RD}\, S_R S_U + E_{z,LU}\, S_L S_D - E_{z,RU}\, S_R S_D - E_{z,LD}\, S_L S_U}{(S_R - S_L)(S_U - S_D)}$$
$$+ \left[\frac{S_U S_D}{(S_R - S_L)(S_U - S_D)}\right]\left[S_R\left(B_{x,RU} - B_{x,RD}\right) - S_L\left(B_{x,LU} - B_{x,LD}\right)\right] \quad (31)$$

We see that the first terms in eqns. (30) and (31) represent convex combinations of the z-component of the electric fields at the four zone corners that abut the z-edge, i.e. O in Fig. 1. The second terms in eqns. (30) and (31) carry the dissipation. Notice that the contributions from the flux components $F_7^*$ and $G_6^*$ are somewhat different. To obtain a unique z-component of the electric field at the z-edge, we have to combine the upwinded flux components from eqn. (30) and (31) in a judicious fashion. We do that next.

Balsara & Spicer [9] were acutely aware that the combination of the upwinded fluxes at zone edges should be carried out multidimensionally and presented an idea from rotated Riemann solvers (Rumsey, Roe & van Leer [45]) to accomplish that. Another approach by Londrillo & Del Zanna [36] and Gardiner & Stone [27] consists of retaining the maximal dissipation terms from either direction while averaging the non-dissipative parts of the flux components. Thus we should suitably average the first terms from eqns. (30) and (31) while we appropriately combine the second terms from the same equations. (Note that $F_7^*$ is, in fact, the negative of the electric field.) This yields a stable scheme for magnetic field update. But the scheme still lacks sufficient cross-term coupling and so its permissible timestep is halved from the desired timestep. There is, however, a modicum of freedom in how the dissipation terms are to be combined. For example, one could use part of the dissipation terms from the multidimensional LLF fluxes, eqns. (19) and (20), because those dissipation terms also introduce cross-term coupling. That resolved the timestep issues. Thus our final form for the electric field is given by

$$E_z = \frac{\frac{1}{2}(E_{z,LU} + E_{z,RD})(S_R S_U + S_L S_D) - \frac{1}{2}(E_{z,RU} + E_{z,LD})(S_R S_D + S_L S_U)}{(S_R - S_L)(S_U - S_D)}$$
$$- (1-\alpha)\left[\frac{S_R S_L}{(S_R - S_L)(S_U - S_D)}\right]\left[S_U\left(B_{y,RU} - B_{y,LU}\right) - S_D\left(B_{y,RD} - B_{y,LD}\right)\right]$$
$$+ \alpha\, \frac{S}{4}\left(B_{y,RU} - B_{y,LU} + B_{y,RD} - B_{y,LD}\right) \quad (32)$$
$$+ (1-\alpha)\left[\frac{S_U S_D}{(S_R - S_L)(S_U - S_D)}\right]\left[S_R\left(B_{x,RU} - B_{x,RD}\right) - S_L\left(B_{x,LU} - B_{x,LD}\right)\right]$$
$$- \alpha\, \frac{S}{4}\left(B_{x,RU} - B_{x,RD} + B_{x,LU} - B_{x,LD}\right)$$

Here $\alpha$ is a parameter that lies between 0 and 1. Smaller values of $\alpha$ are preferred and we have been able to obtain good results with $\alpha = 0.3$. If the underlying numerical method has a shock detector (Colella & Woodward [17], Balsara *et al.* [7]), the previous references show that it pays to smoothly increase $\alpha$ to a value of 0.5 in the vicinity of strong shocks. The directional biasing that is usually built into shock detectors ensures that the detector increases gradually in a zone as a strong shock approaches it. Eqn. (27) may be thought of as a kind of Hamilton-Jacobi equation and the weight of experience, Kurganov, Noelle and Petrova [34], has been that such equations require a larger amount of cross-term dissipation.

Notice that in the subsonic case all of the four electric fields $E_{z,LU}$, $E_{z,RD}$, $E_{z,RU}$ and $E_{z,LD}$ in eqn. (32) contribute to the edge in question. However, they contribute in balanced pairs. To understand the significance of the balanced pairs of electric fields, consider the supersonic case with the flow propagating supersonically into the first quadrant of Fig. 1. Eqn. (17) then gives us $S_L = 0$ and $S_D = 0$ so that the first term of eqn. (32) becomes $(E_{z,LU} + E_{z,RD})/2$. In light of the first example that was studied after eqn. (17) we realize that this is the correct upwinded part that one should get on averaging the x and y-fluxes which is what we did to obtain eqn. (32) from eqns. (30) and (31). If the flow is propagating supersonically into the third quadrant, we should get the same term, and we do. Similarly, if the flow is propagating supersonically into the second or fourth quadrants, the first term of eqn. (32) gives $(E_{z,RU} + E_{z,LD})/2$. We see, therefore, that eqn. (32) does retrieve the correct limits.

**3.2) Comparing the Present Results with Prior Research**

With the present multidimensional Riemann solver in hand, it is also possible to gain insights into previous treatments for the electric field in numerical MHD with a view to understanding their strengths and developing some perspective on their weaknesses.

Londrillo & Del Zanna [36] consider the electric field that is obtained from averaging four one-dimensional HLLE Riemann solvers and provide an explicit formula for it in their eqn. (56). It is, therefore, interesting to compare our eqn. (32) and the analogous eqn. (56) of Londrillo & Del Zanna [36]. Recasting their eqn. (56) in our notation gives

$$E_z = \frac{E_{z,LD}\ S_R S_U + E_{z,RU}\ S_L S_D - E_{z,LU}\ S_R S_D - E_{z,RD}\ S_L S_U}{(S_R - S_L)(S_U - S_D)}$$
$$+ \frac{S_U S_D}{2(S_U - S_D)}(B_{x,RU} - B_{x,RD} + B_{x,LU} - B_{x,LD}) - \frac{S_R S_L}{2(S_R - S_L)}(B_{y,RU} - B_{y,LU} + B_{y,RD} - B_{y,LD})$$
. (33)

We see that their dissipation terms are of the same magnitude, though not the same form, as the ones obtained here. The upwind terms that their formula would pick out in the supersonic limits are not the ones that we obtain from a careful multidimensional analysis.

The above paragraph examined electric fields that are obtained from HLLE Riemann solvers. Strategies for obtaining upwinded, edge-centered electric fields have also been based on other types of one-dimensional Riemann solvers. We examine those strategies next. Thus, analogous to eqn. (7), let $\mathbf{F}_U^{RS}$ denote the flux vector from a general one-dimensional Riemann solver that is applied to the state vectors $\mathbf{U}_{LU}$ and $\mathbf{U}_{RU}$. Again, analogous to eqn. (8), let $\mathbf{F}_D^{RS}$ denote the flux vector from a general one-dimensional Riemann solver that is applied to the states $\mathbf{U}_{LD}$ and $\mathbf{U}_{RD}$. In analogy with eqn. (9), let us use $\mathbf{G}_R^{RS}$ to denote the flux from the one-dimensional Riemann solver applied to the states $\mathbf{U}_{RD}$ and $\mathbf{U}_{RU}$. Similarly, eqn. (10) motivates us to use $\mathbf{G}_L^{RS}$ to denote the flux from the Riemann solver applied to the states $\mathbf{U}_{LD}$ and $\mathbf{U}_{LU}$. The electric field from Balsara & Spicer [9] or Balsara [5] can then be written as

$$E_z = \frac{1}{4}\left[ -\left(\mathbf{F}_U^{RS}\right)_7 - \left(\mathbf{F}_D^{RS}\right)_7 + \left(\mathbf{G}_R^{RS}\right)_6 + \left(\mathbf{G}_L^{RS}\right)_6 \right]. \tag{34}$$

The numerical subscripts in eqn. (34) denote the component of the flux vector. We see that it lacks the enhanced dissipation from eqn. (32), with the result that it may need more dissipation on some subsonic and transonic problems. However, as long as the underlying one-dimensional Riemann solver retrieves the correct supersonic limit, eqn. (34) will pick out the correct upwinded limits in all the supersonic cases. This explains why it is a strong performer on problems with strong, supersonic shocks.

Londrillo & Del Zanna [36] and Gardiner & Stone [27] have presented other strategies for obtaining the electric field that are still based on the dualism of the flux components and electric fields. Eqn. (39) from Gardiner & Stone [27] and eqns. (41) and (42) of Londrillo & Del Zanna [36] both yield the same form given by

$$E_z = \frac{1}{2}\left[ -\left(\mathbf{F}_U^{RS}\right)_7 - \left(\mathbf{F}_D^{RS}\right)_7 + \left(\mathbf{G}_R^{RS}\right)_6 + \left(\mathbf{G}_L^{RS}\right)_6 \right] - \frac{1}{4}\left( E_{z,RU} + E_{z,LD} + E_{z,LU} + E_{z,RD} \right). \tag{35}$$

Notice that eqn. (35) doubles the dissipation in eqn. (34) in the subsonic and transonic cases, where such a doubling of the dissipation is needed. As a result, eqn. (35) performs well in the subsonic and transonic limits. The deficiency in eqn. (35) shows up in the supersonic limit. Say the flow is propagating supersonically into the first quadrant of Fig. 1. One would then expect that $E_z$ should not be independent of any contribution from the downwind direction. Yet, eqn. (35) picks up a piece given by $E_{z,RU}/4$ which diminishes its ability to perform well on problems with strong, supersonic shocks. One of the algorithms presented in Gardiner & Stone [27] (which they refer to as the $\mathcal{E}_z^o$ algorithm in their paper) replaces the second term in eqn. (35) with $\left( \bar{E}_{z,RU} + \bar{E}_{z,LD} + \bar{E}_{z,LU} + \bar{E}_{z,RD} \right)/4$ where $\bar{E}_{z,RU}$ is the zone-averaged value of the electric field in the first quadrant and a similar notation is applied to all other quadrants. This replacement would further increase the downwind character of their scheme in the supersonic limits.

A one-dimensional HLLE Riemann solver can indeed be slightly dissipative when compared to its alternatives. When one must use a different one-dimensional Riemann solver to assemble an edge-centered electric field, a happy compromise would, therefore, consist of using eqn. (35) for the subsonic and transonic cases and using eqn. (34) in the supersonic cases. This is easily accomplished because the doubling of the dissipation that is inherent in eqn. (35) can be withheld in each of the four contributing one-dimensional Riemann solvers when their Riemann fans become supersonic. (All Riemann solvers do indeed check for their Riemann fans being supersonic because it leads to other computational simplifications that are always exploited.) Consequently, our simple strategy can be implemented post-facto at the end of any one-dimensional MHD Riemann solver. It is tantamount to doubling the dissipation in the last three components of the flux vector when the Riemann fan is subsonic and doing nothing when the Riemann fan is supersonic. The four values of the electric fields that come from such a modified Riemann solver can then be combined as in eqn. (34). This was indeed the strategy that was used in Balsara *et al.* [7] and Balsara [6].

**4) Brief Description of the One-Step, Second Order Accurate, Predictor-Corrector Scheme for Euler and MHD Flow**

The multidimensional Riemann solver presented here is inherently two-dimensional. A three dimensional extension of the same is the topic of future research. Just as a one-dimensional Riemann solver is applied to each face and yields one flux, a two-dimensional Riemann solver is applied to each edge and provides two fluxes in the two directions that are transverse to the direction of that edge. (Similarly, a genuinely three dimensional Riemann solver would be applied at each vertex and would yield three fluxes.) Thus the Riemann solver presented here is applied by visiting each edge and solving the multidimensional Riemann problem at that edge. For MHD, that process directly yields the electric field along that edge, see eqn. (32). At each face one would also desire the fluid flux normal to that face for both Euler and MHD flows. In two dimensions this is obtained by using eqn. (24). Similarly eqn. (25) is used in three dimensions. Recall too that for a second order scheme the evaluation of a face centered flux can be simplified in two and three dimensions in light of the discussion that follows eqn. (24).

The calculation can be structured very economically so that the two-dimensional Riemann solver is applied at each edge producing two upwinded fluxes as outputs. See eqns. (15) and (16) for an example of the fluxes, $F^*$ and $G^*$, that are obtained at the z-edge. Eqn. (24) and its analogue for the y-flux can then be used to assemble the facial fluxes in two dimensions. For a two-dimensional calculation, one only needs to apply the two-dimensional Riemann solver to all of the vertices of the two-dimensional mesh. As a result, application of the multidimensional Riemann solver at each of the vertices only results in one call to the multidimensional Riemann solver per zone in a two-dimensional calculation. In three dimensions, each edge is shared by four zones. Consequently, a single application of the two-dimensional Riemann solver to all of the three different edge directions of a three dimensional mesh is tantamount to making three calls to the two-dimensional Riemann solver per zone.

We now provide a pointwise description of the one-step predictor-corrector scheme that we used in the examples that will be presented in the next three Sections. The Riemann solver is called twice at each edge in the course of a timestep. As a result, the present scheme makes six calls to the multidimensional Riemann solver in the course of updating a zone through one timestep. Each timestep of the scheme consists of the following six functional sub-steps.

**1)** *Reconstruct Conserved Variables*: For each zone-centered variable, apply a limiter in each of the dynamically active directions to obtain the undivided differences in those directions.
**2)** *Reconstruct Magnetic Fields*: For face-centered magnetic field components, apply a limiter in each of the two transverse directions that lie within that face. Use the undivided differences that are produced in those faces to obtain a second order accurate divergence-free reconstruction of the magnetic field within each zone (Balsara [4], [5]). This step is not needed for Euler flows.
**3)** *Predictor Step*: We are now in a position to carry out spatially second order accurate interpolation of variables within a zone. Likewise, for a magnetic field component, we can carry out second order accurate spatial interpolation within the face that it belongs to. Owing to the divergence-free reconstruction we can also interpolate any of the magnetic field components to any location within a zone. Therefore, use that spatial interpolation to produce the four states that go into the two-dimensional Riemann solver that is applied at each edge. Visit each edge and obtain the fluxes at the faces for this predictor step. If this is an MHD calculation, obtain the electric fields at the zone edges for this predictor step.
**4)** *Predicted Time Rates of Update*: Use the fluxes to obtain the time rate of update for all the zone-centered variables. This includes obtaining the time rate of update for the zone-centered divergence-free reconstruction of each magnetic field component. Likewise, use the electric fields to obtain a time rate of update for the magnetic field components within each face.
**5)** *Corrector Step*: The time rates of update from the previous step can now be used to make a corrector step that is second order accurate in space and time. Consequently, use that space-time interpolation to produce the four states that go into the two-dimensional Riemann solver that is applied at each edge. Notice that unlike the predictor step, the states in this corrector step are centered in time. Visit each edge and obtain the fluxes at the faces. Also obtain the electric fields at the edges if this is an MHD calculation.
**6)** *Second Order Accurate Update*: The fluxes and electric fields that are obtained from the previous step are centered in space and time. They can then be used to make a one-stage update that is conservative, divergence-free and second order accurate in space and time. See Balsara *et al.* [7] for an example of such a one-step update.

This completes our description of the timestepping strategy.

The algorithm presented here has been implemented in the author's RIEMANN code. In light of the interest in large multidimensional calculations, the implementation was optimized for two and three-dimensional calculations. We made one implementation for Euler flow and another for MHD flow. On a single Intel Xeon 5500 core operating at 2.4 GHz the Euler and MHD codes with the multidimensional Riemann solvers update 136,500 and 80,000 three-dimensional zones per second respectively. (In two dimensions, the codes update 415,700 and 196,000 zones per second for Euler and MHD flow.) A comparable version of the RIEMANN code (that utilizes one-dimensional HLLE Riemann solvers for the predictor and corrector steps) updates 138,600 and 81,250 three-dimensional zones per second for Euler and MHD flow respectively. The difference in the speeds is very slight. The implementation of the multidimensional Riemann solver efficiently organizes the float-point intensive work, doing more calculations per subroutine call. This permits us to amortize the overhead of each subroutine call in a more expeditious fashion. Subsequent sections show that the small difference in the speeds is handily compensated for by the larger timesteps permitted by the multidimensional Riemann solver. MHD calculations are especially benefited by the use of our genuinely multidimensional Riemann solver because it yields a genuinely multidimensional treatment of the edge-centered electric fields. Subsequent sections present accuracy analysis and test problems for the code described here.

**5) Accuracy Analysis for Euler and MHD Flows**

The schemes presented here can easily achieve second order accuracy for one-dimensional problems. Because this is a paper on multidimensional Riemann solvers, we focus on multidimensional demonstrations of second order accuracy. A suite of such test problems was presented in Balsara *et al.* [7]. We present a couple of interesting two-dimensional tests from that test suite. All the two-dimensional test problems in this section were run with a CFL number of 0.65, though we have verified that they also run stably and without any appreciable change in the accuracy when run with a CFL number of 0.85. To provide a point of comparison, all the simulations were first run with a minmod limiter, then they were run with an MC limiter and lastly they were run with the slopes that can be obtained from the r=3 WENO scheme of Jiang and Shu [33]. In other words, the piecewise parabolic part of the reconstruction that can be obtained from the r=3 WENO reconstruction was not retained and the stencils were centrally biased.

**5.1) Unmagnetized Isentropic Vortex in Two Dimensions**

In this hydrodynamical test problem from Balsara & Shu [8], an isentropic vortex is made to propagate for a time of 10 units along the diagonal of a periodic domain spanning $[-5,5]\times[-5,5]$. The unperturbed flow has unit density, unit pressure and unit velocities in each of the x and y-directions. The gas has a ratio of specific heats given by 1.4. The entropy is given by $S = P/\rho^\gamma$ and density and pressure fluctuations are introduced in such a way as to keep the flow isentropic. The temperature is defined by $T = P/\rho$. A vortex is set up as a fluctuation to the unperturbed flow where the fluctuations are specified by

$$(\delta v_x, \delta v_y) = \frac{\varepsilon}{2\pi} e^{0.5(1-r^2)}(-y, x)$$

$$\delta T = -\frac{(\gamma-1)\varepsilon^2}{8\gamma\pi^2} e^{(1-r^2)}$$

$$\delta S = 0$$

Here we set $\varepsilon = 5$. r is the radius from the origin of the domain and can be written as $r^2 = x^2 + y^2$.

TABLE I

Table I shows the accuracy analysis for the two-dimensional, unmagnetized, isentropic vortex problem using schemes that use the multidimensional Riemann solver presented here. Minmod, MC and r = 3 WENO slopes were used. The errors were measured using the density variable which was compared to the analytical solution.

| Method | Number of zones | $L_1$ error | $L_1$ order | $L_\infty$ error | $L_\infty$ order |
|---|---|---|---|---|---|
| minmod limiter | 64×64 | $8.0130 \times 10^{-3}$ | | $1.5466 \times 10^{-1}$ | |
| | 128×128 | $2.6687 \times 10^{-3}$ | 1.59 | $5.9768 \times 10^{-2}$ | 1.37 |
| | 256×256 | $9.0798 \times 10^{-4}$ | 1.55 | $2.7398 \times 10^{-2}$ | 1.13 |
| | 512×512 | $3.3087 \times 10^{-4}$ | 1.46 | $1.5020 \times 10^{-2}$ | 0.86 |
| MC limiter | 64×64 | $2.3608 \times 10^{-3}$ | | $6.1816 \times 10^{-2}$ | |
| | 128×128 | $5.5141 \times 10^{-4}$ | 2.10 | $2.7894 \times 10^{-2}$ | 1.15 |
| | 256×256 | $1.1895 \times 10^{-4}$ | 2.22 | $6.2342 \times 10^{-3}$ | 2.15 |
| | 512×512 | $2.3152 \times 10^{-5}$ | 2.35 | $1.9041 \times 10^{-3}$ | 1.71 |
| WENO limiter | 64×64 | $1.2598 \times 10^{-3}$ | | $2.3001 \times 10^{-2}$ | |
| | 128×128 | $2.3236 \times 10^{-4}$ | 2.43 | $3.6835 \times 10^{-3}$ | 2.64 |
| | 256×256 | $4.0201 \times 10^{-5}$ | 2.53 | $5.7757 \times 10^{-4}$ | 2.67 |
| | 512×512 | $8.0701 \times 10^{-6}$ | 2.32 | $9.6655 \times 10^{-5}$ | 2.58 |

Table I shows the accuracy analysis for the multidimensional Riemann solver-based schemes presented here. The errors are measured in the $L_1$ and $L_\infty$ norms for the density variable. We see that the scheme which uses the r=3 WENO slopes starts out with an intrinsically smaller error than the scheme which uses the minmod limiter. Furthermore, the scheme with the r=3 WENO slopes reaches its design accuracy immediately, even on very small meshes whereas the scheme with the minmod limiter has not yet reached its design accuracy for the meshes shown. The MC limiter obtains results that are intermediate between the minmod limiter and the WENO limiter. The schemes with the minmod and MC limiters fail to meet their design accuracy in the $L_\infty$ norm, whereas the scheme with the WENO limiter succeeds on that front. This inability of schemes that are based on TVD limiters to meet their design accuracy in the $L_\infty$ norm is closely related to the fact that TVD limiters clip off extrema in the flow. Along with showing that the multidimensional Riemann solver itself meets its desired specifications, the results also show that it is worthwhile to invest in a higher quality reconstruction algorithm. The better reconstruction algorithm only costs ~11% more per timestep, yet the data from Table I shows that it often enables us to get almost an order of magnitude improvement in accuracy on meshes of all possible sizes.

**5.2) Magnetized Isodensity Vortex in Two Dimensions**

This MHD test problem was described in Balsara [5]. As with the previous test problem, the vortex moves along the diagonal of a periodic domain spanning $[-5,5] \times [-5,5]$ for a time of 10 units. The unperturbed flow has unit density, unit pressure and unit velocities in each of the x and y-directions. The unperturbed magnetic field is zero. The gas has a ratio of specific heats given by 5/3. The magnetized vortex can now be specified as a perturbation to the flow given by

$$(\delta v_x, \delta v_y) = \frac{\kappa}{2\pi} e^{0.5(1-r^2)}(-y, x),$$

$$(\delta B_x, \delta B_y) = \frac{\mu}{2\pi} e^{0.5(1-r^2)}(-y, x),$$

$$\delta P = \frac{1}{8\pi}(\frac{\mu}{2\pi})^2 (1-r^2) e^{(1-r^2)} - \frac{1}{2}(\frac{\kappa}{2\pi})^2 e^{(1-r^2)}.$$

Here we set $\kappa = 1$ and $\mu = \sqrt{4\pi}$ which makes the Alfven speed of the vortex equal to its rotational speed. r is the radius from the origin of the domain and can be written as $r^2 = x^2 + y^2$.

TABLE II

Table II shows the accuracy analysis for the two-dimensional, magnetized, isentropic vortex problem using schemes that are based on the multidimensional Riemann solver presented here. Minmod, MC and r = 3 WENO slopes were used. The errors were measured using the x-component of the magnetic field which was compared to the analytical solution.

| Method | Number of zones | $L_1$ error | $L_1$ order | $L_\infty$ error | $L_\infty$ order |
|---|---|---|---|---|---|
| minmod limiter | 64×64 | $1.0355 \times 10^{-2}$ | | $1.8623 \times 10^{-1}$ | |
| | 128×128 | $3.4013 \times 10^{-3}$ | 1.59 | $7.6101 \times 10^{-2}$ | 1.29 |
| | 256×256 | $1.2161 \times 10^{-3}$ | 1.49 | $3.0813 \times 10^{-2}$ | 1.30 |
| | 512×512 | $3.4345 \times 10^{-4}$ | 1.82 | $1.2788 \times 10^{-2}$ | 1.28 |
| MC limiter | 64×64 | $2.8301 \times 10^{-3}$ | | $6.2567 \times 10^{-2}$ | |
| | 128×128 | $7.3086 \times 10^{-4}$ | 1.95 | $1.9733 \times 10^{-2}$ | 1.67 |
| | 256×256 | $1.9048 \times 10^{-4}$ | 1.94 | $6.5885 \times 10^{-3}$ | 1.74 |
| | 512×512 | $4.7592 \times 10^{-5}$ | 2.00 | $2.2375 \times 10^{-3}$ | 1.56 |
| WENO limiter | 64×64 | $2.2342 \times 10^{-3}$ | | $3.1604 \times 10^{-2}$ | |
| | 128×128 | $5.2474 \times 10^{-4}$ | 2.09 | $6.5299 \times 10^{-3}$ | 2.28 |
| | 256×256 | $1.2927 \times 10^{-4}$ | 2.02 | $1.6460 \times 10^{-3}$ | 1.99 |
| | 512×512 | $3.2436 \times 10^{-5}$ | 1.99 | $4.1233 \times 10^{-4}$ | 1.99 |

Table II shows the accuracy analysis for the multidimensional Riemann solver-based schemes presented here. The errors are measured in the $L_1$ and $L_\infty$ norms for the x-component of the magnetic field. As in the previous test, we see that the scheme which uses the r=3 WENO slopes starts out with an intrinsically smaller error than the scheme which uses the minmod limiter. Consistent with our previous finding, the scheme with the WENO slopes reaches its design accuracy immediately, even on very small meshes whereas the scheme with the minmod limiter has not yet reached its design accuracy for the meshes shown. The MC limiter obtains results that are intermediate between the minmod limiter and the WENO limiter. As in the previous example, both the minmod and MC limiters fail to meet their design accuracy in the $L_\infty$ norm. Thus our accuracy analysis for this magnetized test problem reinforces our findings from the previous, unmagnetized test problem. We see here too that the multidimensional Riemann solver meets its design goal. Furthermore, we see that it pays to invest in a better reconstruction strategy.

**6) Multidimensional Test Problems for Euler Flow**

In this Section we present a couple of multidimensional Riemann problems and the double Mach reflection problem. The tests shown here were run with a CFL number of 0.65. The ratio of specific heats was set to 1.4 for all of the problems in this Section.

**6.1) Multidimensional Riemann Problems**

Schulz-Rinne, Collins and Glaz [49] showed the value of using multidimensional Riemann problems for calibrating numerical schemes. Brio, Zakharian and Webb [13] provided explicit values for the initial conditions for some of these Riemann problems and we catalogue them here. The first multidimensional Riemann problem consists of setting

$\rho = 0.5313$,    P=0.4,    $v_x = 0.0$,    $v_y = 0.0$      for x>0, y>0

$\rho = 1.0$,    P=1.0,    $v_x = 0.0$,    $v_y = 0.7276$      for x>0, y<0

$\rho = 1.0$,    P=1.0,    $v_x = 0.7276$,    $v_y = 0.0$      for x<0, y>0

$\rho = 0.8$,    P=1.0,    $v_x = 0.0$,    $v_y = 0.0$      for x<0, y<0

The problem initially starts off as two weak shocks and two slip lines. The problem was run on a 400×400 mesh that spans $[-1,1] \times [-1,1]$ and was stopped at a time of 0.52. The MC limiter was used along with the multidimensional Riemann solver designed here. The density variable at the latest time in this problem is shown in Fig. 5a. The final solution can be interpreted as two Mach reflections and two contact surfaces at the intersections of the four shocks. We see that a very pronounced density valley moves to the intersection point of the four shocks, which is in keeping with expectations.

The next multidimensional Riemann problem consists of setting

$\rho = 1.5$,    P=1.5,    $v_x = 0.0$,    $v_y = 0.0$      for x>0, y>0

$\rho = 0.5323$,    P=0.3,    $v_x = 0.0$,    $v_y = 1.206$      for x>0, y<0

$\rho = 0.5323$,    P=0.3,    $v_x = 1.206$,    $v_y = 0.0$      for x<0, y>0

$\rho = 0.1379$,    P=0.029,    $v_x = 1.206$,    $v_y = 1.206$      for x<0, y<0

The problem results in a double Mach reflection and a shock propagating at 45º to the mesh. The problem was run on a 400×400 mesh that spans $[-1,1] \times [-1,1]$ and stopped at a time of 1.1. The MC limiter was used along with the multidimensional Riemann solver designed here. The density variable at the latest time in this problem is shown in Fig. 5b. We see that the mushroom cap is captured very crisply in this problem owing to the use of the multidimensional Riemann solver.

**6.2) Double Mach Reflection Problem**

This very popular test problem has been catalogued in great detail by Woodward and Colella [53], consequently we do not repeat the specifics of the initial conditions here. It represents a Mach 10 shock front that interacts with an oblique wedge that is put in its path. The wedge makes an angle of $60^o$ with the shock front and is located along most of the lower x-boundary in Fig. 6. The flow results in a self-similar double Mach system of shocks along with an mushroom cap structure. The problem was set up on a 960×240 mesh on a domain that spans $[0,4] \times [0,1]$. Figs. 6a, 6b, 6c and 6d show the density, pressure, x-velocity and y-velocity respectively at a time of 0.2 for part of the domain that spans $[0,3] \times [0,1]$. The WENO limiter was used. We see that all shocks are captured crisply and the roll-up of the Mach stem is also clearly visible.

**7) Multidimensional Test Problems for MHD Flow**

We now present several MHD test problems. The tests include a field loop advection problem, a rotor problem, the Orzag-Tang problem and a three-dimensional blast problem.

**7.1) Field Loop Advection Problem**

The problem was catalogued by Gardiner and Stone [27], hence its specification is not repeated here. The problem consists of a two-dimensional loop of magnetic field with very low magnetic pressure relative to the gas pressure. It is advected along the diagonal of a 128x64 zone mesh that spans the domain $[-1,1] \times [-0.5, 0.5]$. The magnetic pressure inside the loop is constant to begin with and falls abruptly to zero outside the loop. The loop is initially confined to a radius of 0.3. The loop starts off at the center of the domain. Fig. 7 shows the magnitude of the magnetic field after the loop has been advected around the domain once and returned back to the center. The simulation was run with the WENO limiter. The simulation shown in Fig. 7 was run with a CFL number of 0.65, though this application runs stably and without any problems with a CFL number as large as 0.95. We see that there is virtually no diffusion of the loop's boundaries and no oscillations in the magnetic pressure within the loop's interior. We, therefore, conclude that the method presented here has adequate multidimensional dissipation for the transonic advection of magnetic fields.

**7.2) The Rotor Problem**

This problem was first presented in Balsara and Spicer [9]. We present it again because it has been claimed by Fuchs, Mishra and Risebro [26] that on very large meshes this problem crashes due to negative pressures. For that reason, this problem is set up on a two-dimensional unit square $[-.5,.5] \times [-.5,.5]$ using a 1000×1000 zone mesh. It consists of a dense, rapidly spinning cylinder, in the center of an initially stationary, light ambient fluid. The two fluids are threaded by a magnetic field that is uniform to begin with and has a value of 2.5 units along the x-axis. The initial gas pressure is set to 0.5 in both fluids. The ambient fluid has unit density. The rotor has a constant density of 10 units out to a radius of 0.1. Between a radius of 0.1 and 0.106 a linear taper is applied to the density so that the density in the cylinder linearly joins the density in the ambient. The taper is, therefore, spread out over six computational zones and it is advisable to keep that number fixed as the resolution is increased or decreased. The ambient fluid is initially static. The rotor rotates with a uniform angular velocity that extends out to a radius of 0.1. At a radius of 0.1 it has a toroidal velocity of one unit. Between a radius of 0.1 and 0.106 the rotor's toroidal velocity drops linearly in the radial velocity from one unit to zero so that at a radius of 0.106 the velocity blends in with that of the ambient fluid. The ratio of specific heats is taken to be 5/3. The problem was run with a WENO limiter using a CFL number of 0.65. Figs. 8a, 8b, 8c and 8d show the density, pressure, magnitude of the velocity and the magnitude of the magnetic field respectively at a time of 0.29. We see that the pressure is positive throughout the computational domain. We have verified that the pressure remains positive through a time of 0.295, the final time quoted by Fuchs, Mishra and Risebro [26]. The results for this problem, when computed on a very large mesh, are seen to match those in Balsara and Spicer [9]. The degradation in the density variable that was reported in Londrillo and Del Zanna [36] is not seen in these simulations just as they were not seen in the original paper that first presented this problem.

### 7.3) The Orzag-Tang Problem

This well-known problem by Orzag and Tang [38] was initialized on a periodic domain spanning $[0,2] \times [0,2]$ with the following parameters

$$(\rho,\ v_x,\ v_y,\ P,\ B_x,\ B_y) = (\gamma^2,\ -\gamma \sin(\pi y),\ \gamma \sin(\pi x),\ \gamma,\ -\sin(\pi y),\ \sin(2\pi x)).$$

Here we used $\gamma = 5/3$. The problem was run to a final time of unity with a CFL number of 0.6. The simulation used the MC limiter. To ameliorate concerns about pressure positivity we used a large mesh with 1000×1000 zones. Figs. 9a, 9b, 9c and 9d show the final density, pressure, magnitude of the velocity and the magnitude of the magnetic field respectively. We see that the density and pressure have remained positive. The simulation forms a current sheet with oppositely oriented x-components of magnetic field in the center of the computational domain, as can be surmised from Fig. 9d. The velocity field also shows fluid squirting out in the positive and negative x-directions at the location of the current sheet.

### 7.4) The 3d Blast Problem

A version of this problem was described in Balsara *et al.* [7]. This problem was initialized on the three-dimensional unit domain $[-.5,.5]\times[-.5,.5]\times[-.5,.5]$ using a $129^3$ zone mesh. A unit density was initialized all over the computational domain. The pressure was set to 0.1 all over except within a central sphere of radius 0.1 where it was set to 1000. A magnetic field with a magnitude of 40 was initialized along the (1,1,1) diagonal of the computational domain. The problem was run with a CFL number of 0.4 to a final time of 0.013. A WENO limiter was used. In this three-dimensional problem, eqns. (15) and (25) along with their analogues in the other directions were used for the flux calculations while eqn. (32) and its analogues in the other directions were used for evaluating the electric fields. The variables in the central xy-plane are shown in Fig. 10. Figs. 10a, 10b, 10c and 10d show the final density, pressure, magnitude of the velocity and the magnitude of the magnetic field respectively. Owing to the very large central pressure at the start of the calculation, an extremely strong blast wave propagates outwards, leaving a very low density region in the center. The magnetic field is strongly compressed as the blast wave propagates outwards. Despite the very low plasma beta and the presence of strong shocks, we see that the density and pressure have remained positive in our simulation. The correct multidimensional upwinding that is achieved by the Riemann solver in the supersonic limit plays a very important role in simulating this problem correctly. We have not seen similarly stringent problems performed with several other codes, including those that claimed to use multidimensionally upwinded electric fields.

**8) Conclusions**

We have presented a multidimensional version of the HLLE Riemann solver. This has been accomplished via a simple constructive strategy which introduces one constant resolved state between the states being considered. The introduction of the constant state also introduces adequate dissipation for all equations that are in conservation form. Closed form expressions for the resolved fluxes are also provided to facilitate numerical implementation. In the Appendix, the Riemann solver is shown to be positively conservative when divergence of the fluid velocities is restricted so as to prevent the formation of cavitation in the flow. We also apply our method to obtain multidimensionally upwinded electric fields for divergence-free formulations of the MHD equations.

An efficient second order scheme for Euler and MHD flows that is based on our multidimensional Riemann solver is also presented here and has been implemented in the author's RIEMANN code. It is shown to be cost-competitive with schemes that are based on one-dimensional Riemann solvers while permitting larger timesteps. Accuracy analysis for multidimensional Euler and MHD problems shows that the scheme meets its design accuracy. Several stringent test problems involving Euler and MHD flows are also presented and the scheme is shown to perform robustly on all of them.

**Acknowledgements**

DSB acknowledges support via NSF grant AST-0607731. DSB also acknowledges NASA grants NASA-NNX07AG93G and NASA-NNX08AG69G. The majority of simulations were performed on PC clusters at UND but a few initial simulations were also performed at NASA-NCCS. It is a pleasure to thank M. Dumbser for a very helpful reading of the manuscript.

**Appendix**

Einfeldt *et al.* [22] showed that all higher order Godunov schemes obtain their updated values from a convex averaging process, applied to the states that occur in the solution of the Riemann problem. Consequently, a Riemann solver leads to a positively conservative scheme if and only if all the states generated by the Riemann solver are physically real. In practice this means that if the left and right states of a Riemann solver are physical, i.e. known not to produce a cavitation, then the Riemann solver should also produce a physical resolved state, i.e. one without negative densities or pressures. While this is exactly true when face-centered fluxes are used, the facial fluxes for the scheme described in Section 4 are also obtained through a convex averaging process applied to the fluxes that come from the edge-centered multidimensional Riemann solver. As a result, we can claim that a multidimensional Riemann solver is positively conservative if four physically real states in the four quadrants of Fig. 1 yield a resolved state that has positive density and pressure. As shown in Section 2, when all the variations are restricted to one dimension, the present Riemann solver reduces to the HLLE Riemann solver. As a result, for one-dimensional flows, the present multidimensional HLLE Riemann solver is provably positively conservative.

Let us now focus on proving the positivity of the density variable in our multidimensional HLLE Riemann solver for Euler and MHD flow. The proof extends to all situations where a density-like variable $\rho$ has x and y-fluxes given by $\rho v_x$ and $\rho v_y$ respectively. In other words, our proof of the positivity of the density variable might even encompass relativistic hydrodynamics and relativistic MHD. For Euler flow, the signal speed for the propagation of fluctuations is isotropic in the fluid's rest frame, for MHD it is not. We want our proof of the positivity of the density variable to encompass both cases. As a result, we assume that for a uniform slab of fluid or magnetofluid, the extremal speeds in the x-direction are $v_x - c_x$ and $v_x + c_x$. Similarly, we assume that the extremal speeds in the y-direction are $v_y - c_y$ and $v_y + c_y$. For Euler flow, $c_x = c_y = c$ where "c" is the sound speed. For MHD flow, $c_x$ and $c_y$ are the fast magnetosonic speeds in the x and y-directions. We also restrict attention to the subsonic case, i.e. $S_R S_L < 0$ and $S_U S_D < 0$. We can use eqn. (4) to write the resolved state density, $\rho^*$, as

$$\rho^* = \frac{1}{(S_R - S_L)(S_U - S_D)} \begin{cases} \rho_{RU}\left[(S_R - v_{x,RU})(S_U - v_{y,RU}) - v_{x,RU} v_{y,RU}\right] \\ +\rho_{LD}\left[(S_L - v_{x,LD})(S_D - v_{y,LD}) - v_{x,LD} v_{y,LD}\right] \\ +\rho_{RD}\left[-(S_R - v_{x,RD})(S_D - v_{y,RD}) + v_{x,RD} v_{y,RD}\right] \\ +\rho_{LU}\left[-(S_L - v_{x,LU})(S_U - v_{y,LU}) + v_{x,LU} v_{y,LU}\right] \end{cases} . \quad \text{(A.1)}$$

Note that the curly bracket in eqn. (A.1) does not denote a matrix equation; it is just a sum of four terms. In what follows we show that $\rho^*$ is a convex combination of $\rho_{RU}$, $\rho_{LD}$, $\rho_{RD}$ and $\rho_{LU}$ in the subsonic case. To accomplish this, we have to show that the terms inside the square brackets in eqn. (A.1) are positive. For notational simplicity we define

$$a_{RU} \equiv (S_R - v_{x,RU})(S_U - v_{y,RU}) - v_{x,RU} v_{y,RU}$$
$$a_{LD} \equiv (S_L - v_{x,LD})(S_D - v_{y,LD}) - v_{x,LD} v_{y,LD}$$
$$a_{RD} \equiv -(S_R - v_{x,RD})(S_D - v_{y,RD}) + v_{x,RD} v_{y,RD}$$
$$a_{LU} \equiv -(S_L - v_{x,LU})(S_U - v_{y,LU}) + v_{x,LU} v_{y,LU}$$

(A.2)

which allows us to rewrite eqn. (A.1) compactly as

$$(S_R - S_L)(S_U - S_D)\rho^* = a_{RU}\rho_{RU} + a_{LD}\rho_{LD} + a_{RD}\rho_{RD} + a_{LU}\rho_{LU} \tag{A.3}$$

In the next paragraph we will prove that the first square bracket in eqn. (A.1), i.e. the term $a_{RU}$ from eqn. (A.2), is always positive. Similar arguments can be used to show that the other three square brackets in eqn. (A.1) are also positive.

In the subsonic case we have

$$S_R \geq v_{x,RU} + c_{x,RU} \Rightarrow S_R - v_{x,RU} \geq c_{x,RU} > 0, \tag{A.4}$$

$$S_U \geq v_{y,RU} + c_{y,RU} \Rightarrow S_U - v_{y,RU} \geq c_{y,RU} > 0, \tag{A.5}$$

$$v_{x,RU} \in (-c_{x,RU}, c_{x,RU}) \text{ and } v_{y,RU} \in (-c_{y,RU}, c_{y,RU}) \Rightarrow c_{x,RU} c_{y,RU} > v_{x,RU} v_{y,RU}.$$

(A.6)

The above three conditions allow us to claim that

$$(S_R - v_{x,RU})(S_U - v_{y,RU}) \geq c_{x,RU} c_{y,RU} > v_{x,RU} v_{y,RU} \tag{A.7}$$

which establishes the positivity of the coefficient of $\rho_{RU}$ in eqn. (A.1). Similar arguments can be designed for establishing the positivity of the three other square brackets in eqn. (A.1). This proves that the density of the resolved state, $\rho^*$, is positive definite if the four densities $\rho_{RU}$, $\rho_{LD}$, $\rho_{RD}$ and $\rho_{LU}$ are all positive definite.

We are now in a position to obtain the momentum density and energy density in the resolved state. Using eqn. (4) for the second and third components of the Euler equation gives us

$$(S_R - S_L)(S_U - S_D)\rho^* v_x^* = a_{RU}\rho_{RU} v_{x,RU} + a_{LD}\rho_{LD} v_{x,LD} + a_{RD}\rho_{RD} v_{x,RD} + a_{LU}\rho_{LU} v_{x,LU}$$
$$- S_U(P_{RU} - P_{LU}) + S_D(P_{RD} - P_{LD}) \quad \text{(A.8)}$$

and

$$(S_R - S_L)(S_U - S_D)\rho^* v_y^* = a_{RU}\rho_{RU} v_{y,RU} + a_{LD}\rho_{LD} v_{y,LD} + a_{RD}\rho_{RD} v_{y,RD} + a_{LU}\rho_{LU} v_{y,LU}$$
$$- S_R(P_{RU} - P_{RD}) + S_L(P_{LU} - P_{LD}). \quad \text{(A.9)}$$

Eqns. (A.3), (A.8) and (A.9) can be used to obtain closed form expressions for the velocities $v_x^*$ and $v_y^*$ in the resolved state. These velocities would be useful in formulating an HLLC-type Riemann solver. The resolved state energy density is given by the following equation

$$(S_R - S_L)(S_U - S_D)\mathcal{E}^* = a_{RU}\mathcal{E}_{RU} + a_{LD}\mathcal{E}_{LD} + a_{RD}\mathcal{E}_{RD} + a_{LU}\mathcal{E}_{LU}$$
$$- S_U(P_{RU} v_{x,RU} - P_{LU} v_{x,LU}) + S_D(P_{RD} v_{x,RD} - P_{LD} v_{x,LD}) \quad \text{(A.10)}$$
$$- S_R(P_{RU} v_{y,RU} - P_{RD} v_{y,RD}) + S_L(P_{LU} v_{y,LU} - P_{LD} v_{y,LD})$$

We now present a constructive procedure for demonstrating pressure positivity for the Euler equations in two dimensions. Realize, however, that in their proof of the pressure positivity for the one-dimensional HLLE Riemann solver, Einfeldt *et al.* [22] were helped by the fact that they knew the exact range of diverging velocities that can be sustained in a one-dimensional fluid flow before the fluid undergoes a physical cavitation. In other words, they knew that a physical cavitation occurs when the velocities are divergent at a zone boundary and the Mach number on either side exceeds $2/(\gamma-1)$. The multidimensional Riemann problem has been studied by Schulz-Rinne, Collins and Glaz [49] but they do not arrive at closed form conditions for the velocity ranges that would prevent a cavitation from forming in the flow. As a result, we cannot identify the physically realizable states that would prevent the formation of a cavitation. This is the first difficulty that one encounters in demonstrating pressure positivity in multidimensions. While the mathematics built up here parallels that in Einfeldt *et al.* [22] and reduces to it for one-dimensional variations, the expressions obtained here are much more complicated. As a result, we have to manipulate them numerically on a computer. This precludes closed form expressions with a clear analytical proof which is, consequently, our second difficulty in demonstrating pressure positivity. Our procedure for showing pressure positivity is, therefore, a little less than an ironclad proof.

Demonstrating the pressure positivity of the resolved state is tantamount to showing that the term

$$(S_R - S_L)^2 (S_U - S_D)^2 \left[\rho^* \mathcal{E}^* - \frac{1}{2}(\rho^* v_x^*)^2 - \frac{1}{2}(\rho^* v_y^*)^2\right] \quad \text{(A.11)}$$

is always positive. After much algebra, this term can be written as

$$(S_R - S_L)^2 (S_U - S_D)^2 \left[ \rho^* \varepsilon^* - \frac{1}{2}(\rho^* v_x^*)^2 - \frac{1}{2}(\rho^* v_y^*)^2 \right] =$$

$$\frac{1}{\gamma - 1}\left(a_{RU}\rho_{RU} + a_{LD}\rho_{LD} + a_{RD}\rho_{RD} + a_{LU}\rho_{LU}\right)\left(a_{RU}P_{RU} + a_{LD}P_{LD} + a_{RD}P_{RD} + a_{LU}P_{LU}\right)$$

$$+ \frac{1}{2} a_{RU}\rho_{RU} a_{LU}\rho_{LU} \left[ (v_{x,RU} - v_{x,LU})^2 + (v_{y,RU} - v_{y,LU})^2 \right] + \frac{1}{2} a_{RU}\rho_{RU} a_{RD}\rho_{RD} \left[ (v_{x,RU} - v_{x,RD})^2 + (v_{y,RU} - v_{y,RD})^2 \right]$$

$$+ \frac{1}{2} a_{RU}\rho_{RU} a_{LD}\rho_{LD} \left[ (v_{x,RU} - v_{x,LD})^2 + (v_{y,RU} - v_{y,LD})^2 \right] + \frac{1}{2} a_{LU}\rho_{LU} a_{LD}\rho_{LD} \left[ (v_{x,LU} - v_{x,LD})^2 + (v_{y,LU} - v_{y,LD})^2 \right]$$

$$+ \frac{1}{2} a_{LU}\rho_{LU} a_{RD}\rho_{RD} \left[ (v_{x,LU} - v_{x,RD})^2 + (v_{y,LU} - v_{y,RD})^2 \right] + \frac{1}{2} a_{LD}\rho_{LD} a_{RD}\rho_{RD} \left[ (v_{x,LD} - v_{x,RD})^2 + (v_{y,LD} - v_{y,RD})^2 \right]$$

$$+ \left(S_U P_{RU} + S_D P_{LD} - S_D P_{RD} - S_U P_{LU}\right)\left(a_{RU}\rho_{RU}v_{x,RU} + a_{LD}\rho_{LD}v_{x,LD} + a_{RD}\rho_{RD}v_{x,RD} + a_{LU}\rho_{LU}v_{x,LU}\right)$$

$$+ \left(S_R P_{RU} + S_L P_{LD} - S_R P_{RD} - S_L P_{LU}\right)\left(a_{RU}\rho_{RU}v_{y,RU} + a_{LD}\rho_{LD}v_{y,LD} + a_{RD}\rho_{RD}v_{y,RD} + a_{LU}\rho_{LU}v_{y,LU}\right)$$

$$- \left(a_{RU}\rho_{RU} + a_{LD}\rho_{LD} + a_{RD}\rho_{RD} + a_{LU}\rho_{LU}\right)\left(S_U P_{RU} v_{x,RU} + S_D P_{LD} v_{x,LD} - S_D P_{RD} v_{x,RD} - S_U P_{LU} v_{x,LU}\right)$$

$$- \left(a_{RU}\rho_{RU} + a_{LD}\rho_{LD} + a_{RD}\rho_{RD} + a_{LU}\rho_{LU}\right)\left(S_R P_{RU} v_{y,RU} + S_L P_{LD} v_{y,LD} - S_R P_{RD} v_{y,RD} - S_L P_{LU} v_{y,LU}\right)$$

$$- \frac{1}{2}\left[ -S_U (P_{RU} - P_{LU}) + S_D (P_{RD} - P_{LD}) \right]^2 - \frac{1}{2}\left[ -S_R (P_{RU} - P_{RD}) + S_L (P_{LU} - P_{LD}) \right]^2$$

. (A.12)

Eqn. (A.12) is exactly analogous to eqn. (A8) from Einfeldt *et al*. [22] and for one-dimensional variations our eqn. (A.12) indeed reduces to their eqn. (A8). As a result, at least for one-dimensional variations, the multidimensional HLLE Riemann solver retains pressure positivity on an equal footing with the one-dimensional HLLE Riemann solver. Notice that eqn. (A.12) is, at least formally, a quadratic function in the velocities. The subsonic condition ensures that the coefficients in front of the quadratic velocity terms in eqn. (A.12) are all positive definite. If we think of the densities and pressures in the four quadrants of Fig. 1 as being specified, we can then view the right hand side of eqn. (A.12) as a function in eight variables formed by the velocities $v_{x,RU}, v_{x,LD}, v_{x,RD}, v_{x,LU}$ and $v_{y,RU}, v_{y,LD}, v_{y,RD}, v_{y,LU}$. Denote the right hand side of eqn. (A.12) by the function $\psi(v_{x,RU}, v_{x,LD}, v_{x,RD}, v_{x,LU}, v_{y,RU}, v_{y,LD}, v_{y,RD}, v_{y,LU})$ As a result, the function $\psi$ defines a paraboloid with positive quadratic terms in the velocity variables. We need to show that the minimum of $\psi$ is also positive when all the four input states from the four quadrants in Fig. 1 are physical. Because $\psi$ is, at least formally, quadratic in the velocities. We can find the minimum of $\psi$ by setting

$$\frac{\partial \psi}{\partial v_{x,RU}} = \frac{\partial \psi}{\partial v_{x,LD}} = \frac{\partial \psi}{\partial v_{x,RD}} = \frac{\partial \psi}{\partial v_{x,LU}} = \frac{\partial \psi}{\partial v_{y,RU}} = \frac{\partial \psi}{\partial v_{y,LD}} = \frac{\partial \psi}{\partial v_{y,RD}} = \frac{\partial \psi}{\partial v_{y,LU}} = 0. \quad (A.13)$$

Notice that for any specified set of densities and pressures in the four quadrants of Fig. 1, eqn. (A.12) is only a quadratic at a formal level. Thus eqn. (A.13) will only permit us to iterate to the set of velocities that minimize $\psi$. In practice, as the velocities change, so do $S_R$, $S_L$, $S_U$ and $S_D$. Consequently, the coefficients in eqn. (A.2) also change. These changes are, however, made with the proviso that the subsonic condition is met, thus keeping the coefficients in eqn. (A.2) positive. By iterating eqn. (A.13) to convergence for any specified set of densities and pressures in the four quadrants of Fig. 1 and evaluating $\psi$ for that minimum point, we can prove the pressure positivity for *all* possible velocities. This discussion suggests that the procedure is very amenable to implementation on a computer.

There are several subtler points that need to be kept in mind for a good computer implementation of a demonstration (not proof) of pressure positivity. First, realize that the Euler equations are Galilean invariant. As a result, eqn. (A.13) yields a degenerate system. All we can do is to specify $v_{x,RU}$ and $v_{y,RU}$ for one of the four slabs of fluid and use the remaining six equations in eqn. (A.13) to obtain the other velocities relative to the specified ones. For the sake of simplicity, we set $v_{x,RU} = v_{y,RU} = 0$. Eqn. (A.13) is a Newton step in a root solver with six free variables and, therefore, we have always observed reasonably rapid convergence. Also notice that the Euler equations are scale invariant. As a result, we can set $\rho_{RU} = P_{RU} = 1$ for one of the four slabs of fluid. The densities and pressures in the other three slabs of fluids are then specified as ratios relative to $\rho_{RU}$ and $P_{RU}$. Thus the independent variables are the six ratios $\rho_{LD}/\rho_{RU}$, $P_{LD}/P_{RU}$, $\rho_{RD}/\rho_{RU}$, $P_{RD}/P_{RU}$, $\rho_{LU}/\rho_{RU}$ and $P_{LU}/P_{RU}$. In the computer code, we have a six-fold nested loop that causes these ratios to independently span the range $[\,0.1\,,\,102.4\,]$. To span this large range efficiently, we use geometric scaling so that each loop iteration increases the ratio of the variable it governs by a factor of 2 relative to the previous iteration. For each iteration of the sextuply nested loop we can find the velocities that minimize $\psi$.

Two further subtleties still need to be mentioned. First, notice that if the velocities are unconstrained, the physics of the problem is such that $\psi$ can always attain a negative value if the flows in all the four quadrants diverge supersonically from the edge being considered. The computer code for minimizing $\psi$ indeed gravitates to such solutions in several cases that we tested. (When variations are restricted to one dimension, we were able to verify that the computer code retrieves the full acceptable range of velocities from Einfeldt *et al.* [22].) Because we wish to show pressure positivity in the subsonic case, the velocities in the four quadrants have to be restricted to be subsonic. We, therefore, need to limit the range of velocities being considered. This effort to restrict velocities to some sort of subsonic range reminds us of our second difficulty. Realize that in multidimensions we do not have closed form expressions for ranges of velocities that do not result in a cavitation. Recall that the proof of pressure positivity in Einfeldt *et al.* [22] was only possible because they were first able to bound the range of velocities so as to exclude cavitations. For the multidimensional case, we have no a priori knowledge of what that range should be. We, therefore, make the reasonable assumption that:

$$\text{if } (v_{x,LD} < 0) \, v_{x,LD} = -\min\left(\left|v_{x,LD}\right|, \phi\, c_{LD}, \phi\, c_{RD}, \phi\, c_{RU}\right)$$

$$\text{if } (v_{y,LD} < 0) \, v_{y,LD} = -\min\left(\left|v_{y,LD}\right|, \phi\, c_{LD}, \phi\, c_{LU}, \phi\, c_{RU}\right)$$

$$\text{if } (v_{x,RD} > 0) \, v_{x,RD} = \min\left(\left|v_{x,RD}\right|, \phi\, c_{RD}, \phi\, c_{LD}, \phi\, c_{LU}\right)$$

$$\text{if } (v_{y,RD} < 0) \, v_{y,RD} = -\min\left(\left|v_{y,RD}\right|, \phi\, c_{RD}, \phi\, c_{RU}, \phi\, c_{LU}\right) \quad (A.14)$$

$$\text{if } (v_{x,LU} < 0) \, v_{x,LU} = -\min\left(\left|v_{x,LU}\right|, \phi\, c_{LU}, \phi\, c_{RU}, \phi\, c_{RD}\right)$$

$$\text{if } (v_{y,LU} > 0) \, v_{y,LU} = \min\left(\left|v_{y,LU}\right|, \phi\, c_{LU}, \phi\, c_{LD}, \phi\, c_{RD}\right)$$

Here $c_{RU}$ denotes the sound speed of the fluid in the first quadrant, with a similar notation extending to the other three slabs of fluid. Notice that $\phi$ is positive, so the six conditions in eqn. (A.14) make no restrictions on situations where slabs of fluids collide with each other. The above six conditions only restrict how fast slabs of fluid recede from each other, resulting in a multidimensional rarefaction. In other words, $\phi$ is a measure of the divergence of the multidimensional flow velocity. The condition in eqn. (A.14) only restricts large positive values of the divergence in the flow velocity. As a result, larger values of $\phi$ permit the undivided divergence in the flow velocity to reach larger positive values. Since these velocities are scaled by the local sound speed in eqn. (A.14), $\phi$ also controls the Mach number of the undivided divergence of the flow. Consequently, just as Einfeldt *et al*. [22] had to restrict their one-dimensional undivided divergence in the flow velocity to lie within a certain range of Mach numbers, our use of $\phi$ enables us to make a similar restriction in multiple dimensions.

Using the computer code we wrote for demonstrating pressure positivity, we were able to guarantee that with $\phi \leq 0.43$ the pressures remain positive when the six ratios $\rho_{LD}/\rho_{RU}$, $P_{LD}/P_{RU}$, $\rho_{RD}/\rho_{RU}$, $P_{RD}/P_{RU}$, $\rho_{LU}/\rho_{RU}$ and $P_{LU}/P_{RU}$ independently span the range $[0.1, 102.4]$. Admittedly, setting $\phi \leq 0.43$ produces a range of diverging velocities that is smaller than the corresponding range of divergent velocities that is allowed in one-dimensional flow. However, we do not have an available theory for two-dimensional rarefaction flows that we can draw on. The range that we have demonstrated is still quite good for practical work. We have experimented with larger values of $\phi$ and find that all the situations where pressure positivity is relinquished occur when the computer code finds flow velocities in all the quadrants that are diverging away from the edge being considered, i.e. multidimensional rarefactions. We have no theory to tell us whether those situations are physically realizable of not. When $\phi = 0.6$ was handed in to the code, we found that only 0.1% of the cases tested yielded negative pressures. When $\phi = 0.75$ and $0.9$ were handed in to the code, we found that only 0.52% and 1.15% respectively of the cases tested yielded negative pressures. Because of the heuristic bounds given in eqn. (A.14) please note that we have no criterion for telling whether the very divergent velocities in the code are physically realizable or not. Even so, when very divergent velocities are allowed, the cases where pressure positivity is lost are indeed very small in number. The code for demonstrating pressure positivity is available by emailing the author from a bona fide academic email address.

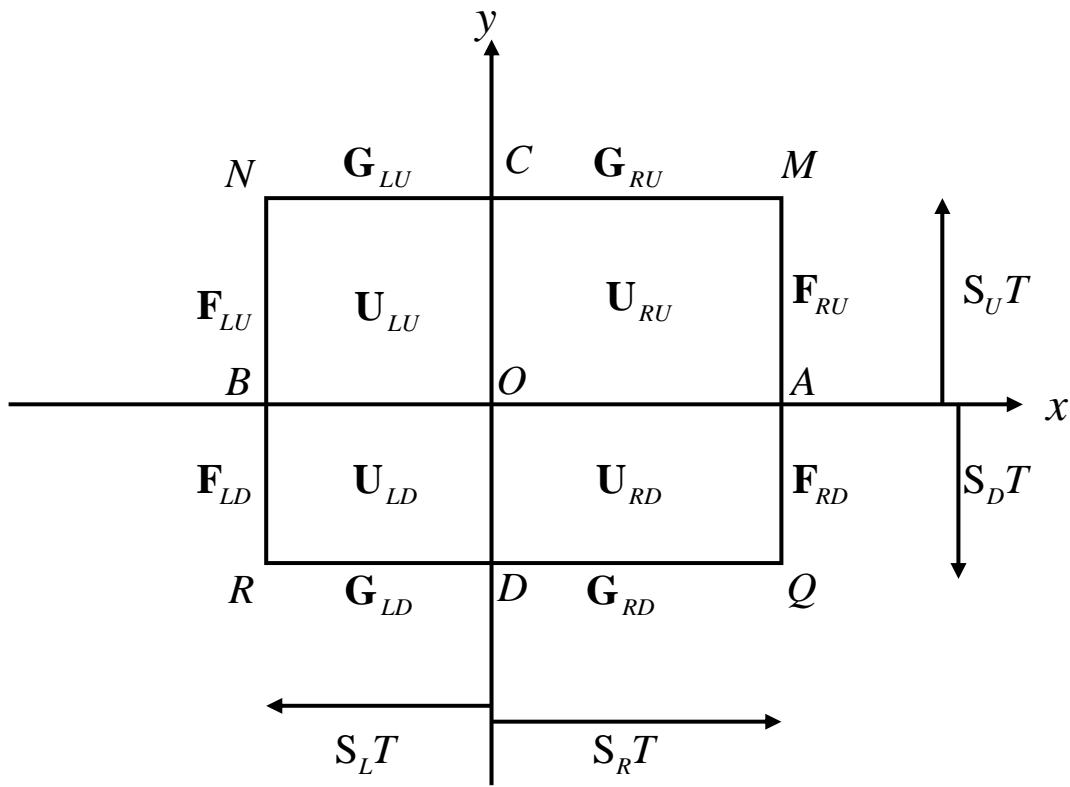

Fig. 1 depicts a situation where four neighboring zones meet at an edge. (Think of the zones as having an extension in the third dimension.) The four zones lie in each of the four quadrants of the xy-plane. The origin O of the xy-plane denotes the edge shared by the four zones. The solution vector and fluxes in the first quadrant are denoted by a subscript RU (right-up); those in the second quadrant are shown by a subscript LU (left-up); those in the third quadrant have a subscript LD (left-down); those in the fourth quadrant carry a subscript RD (right-down). The waves start propagating outward from the origin at $t=0$. In a time $t=T$, the waves propagate out to $x = S_R T$ and $x = S_L T$ along the x-axis and out to $y = S_U T$ and $y = S_D T$ along the y-axis. The rectangle QMNR bounds the domain that will be affected by the waves.

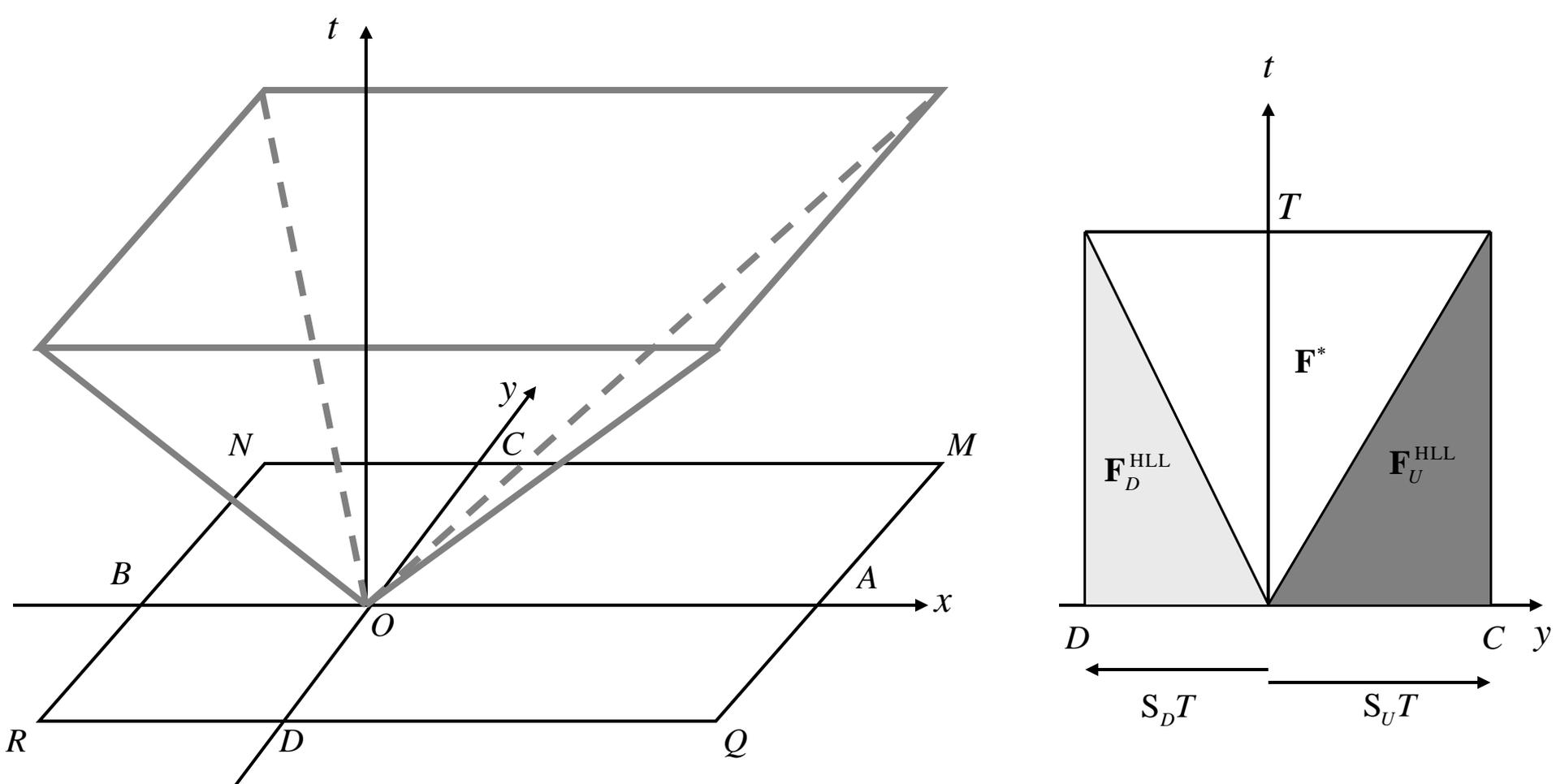

*Fig. 2. The left panel shows the wave model that we adopt for the propagation of waves in two space dimensions and one temporal dimension. Points in space-time that are contained within the inverted, dark gray, rectangular pyramid in this figure are within the range of influence of the initial discontinuity. The wave model circumscribes the actual waves propagating out of the initial discontinuity at O. The right panel shows the plane x=0 from the left panel along with the x-directional fluxes that propagate through different portions of that face. Thus the resolved flux $\mathbf{F}^*$ propagates through the unshaded area; the flux $\mathbf{F}_D^{\mathrm{HLL}}$ propagates through the light gray area and the flux $\mathbf{F}_U^{\mathrm{HLL}}$ propagates through the dark gray area.*

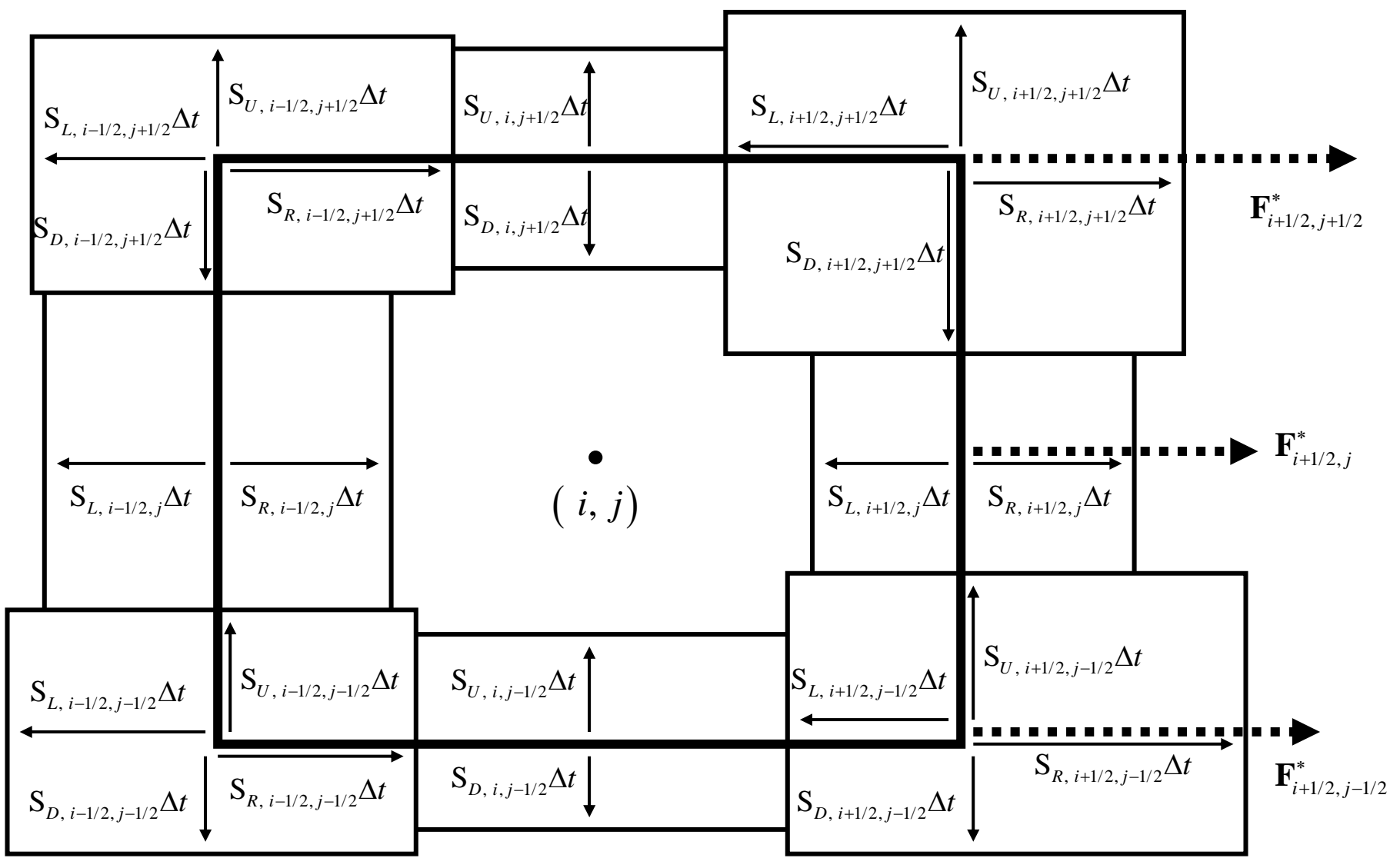

Fig. 3 shows the evolution of Riemann problems at all the faces and all the corners of a two-dimensional zone for a time $\Delta t$. As time $\Delta t$ increases, the multidimensional Riemann problems at each of the corners make an increasingly larger contribution to the facially and temporally averaged fluxes at the zone faces. This is especially true in the subsonic cases shown here. The solid arrows in this figure show the propagation of waves; the dashed arrows show the fluxes.

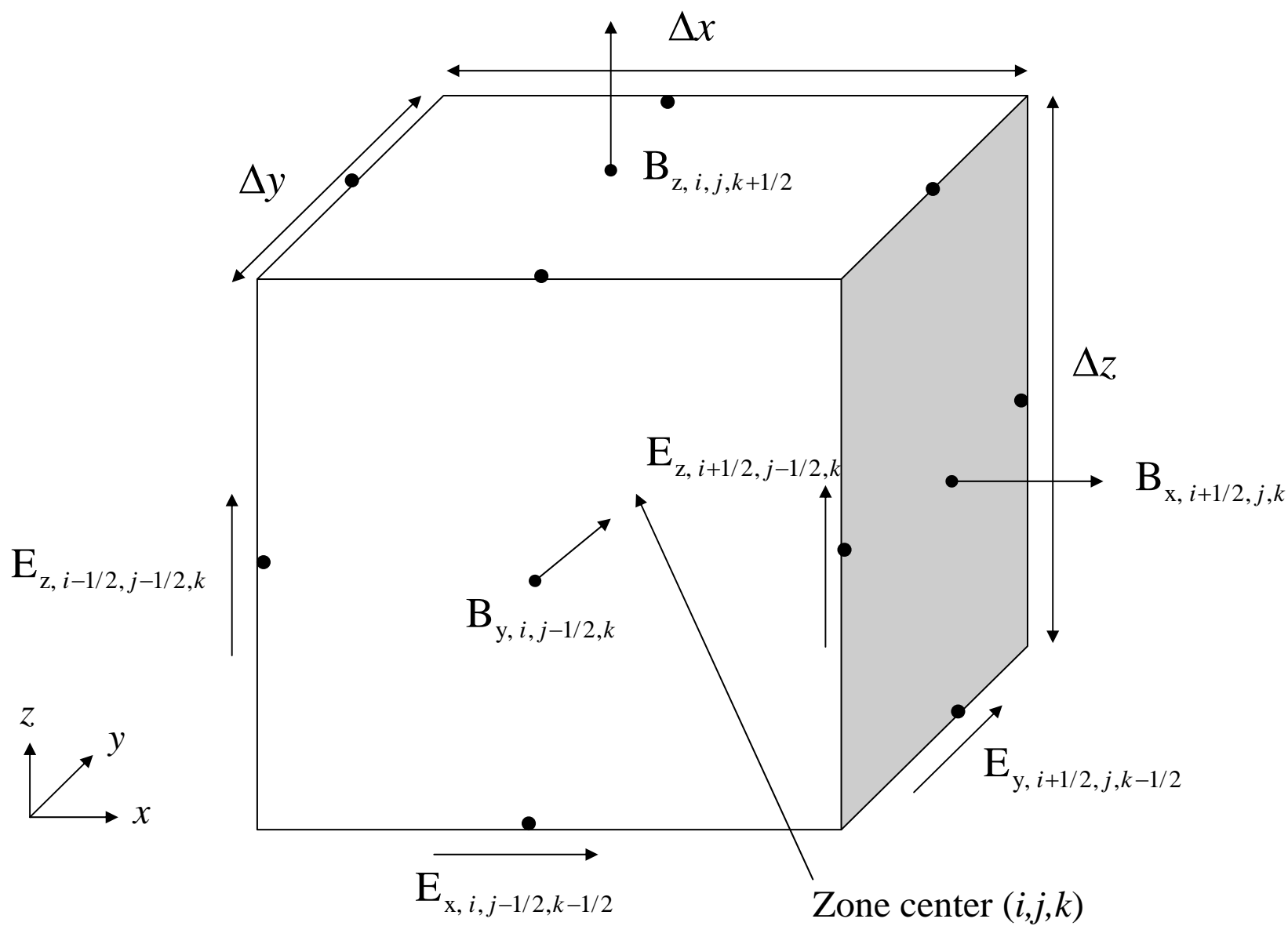

*Fig. 4 showing the collocation of face-centered magnetic fields and edge-centered electric fields in a constrained transport method.*

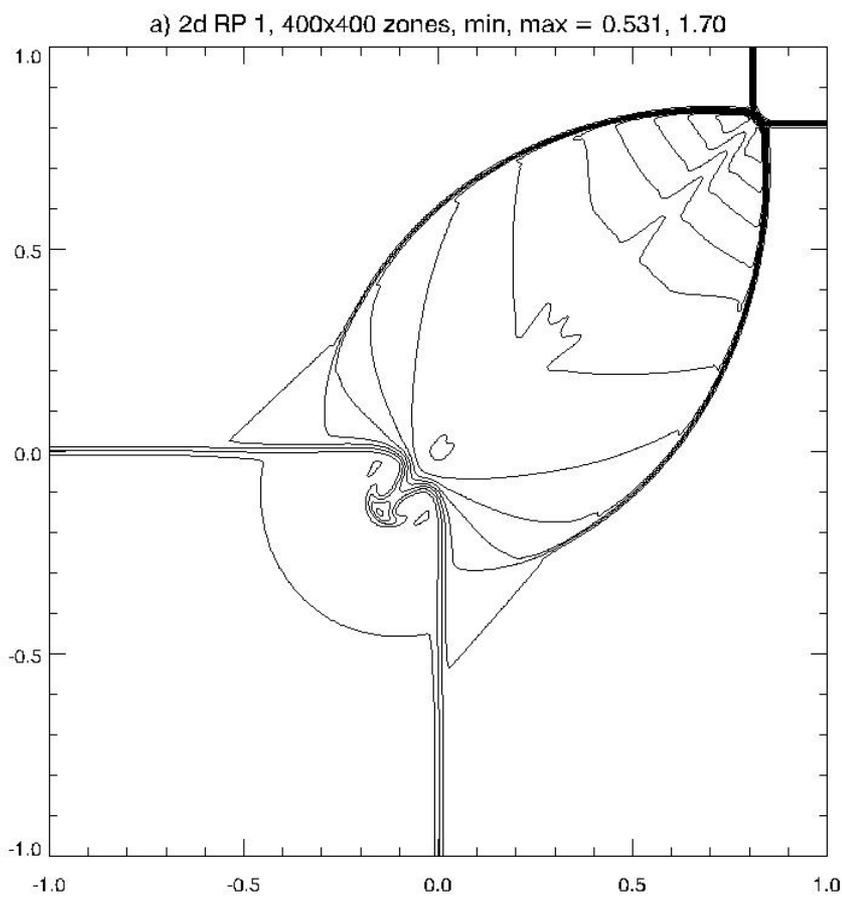 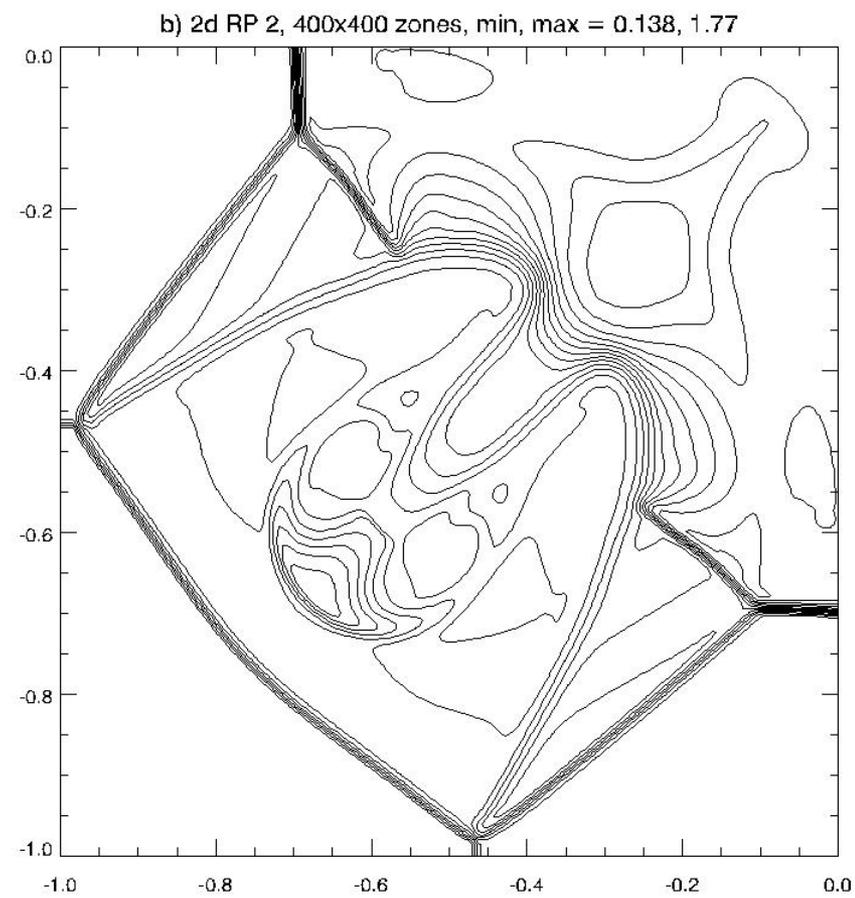

*Figs. 5a and 5b show contour plots of the final densities from the first and second multidimensional Riemann problems at the latest times in those simulations. The whole domain had $400^2$ zones. 20 contours are displayed for both figures. In Fig. 5a, the contours span [0.531, 1.70]. In Fig. 5b they span the range [0.138,1.77]. Fig. 5b only shows the lower left quadrant of the computational domain since the rest of the computational domain does not develop any interesting structures in the fluid variables.*

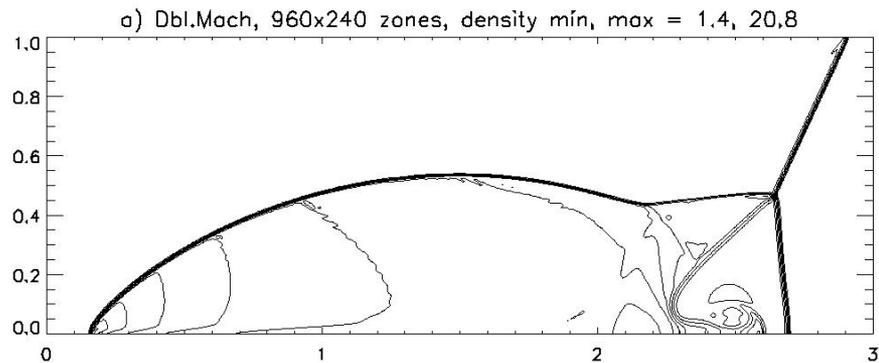
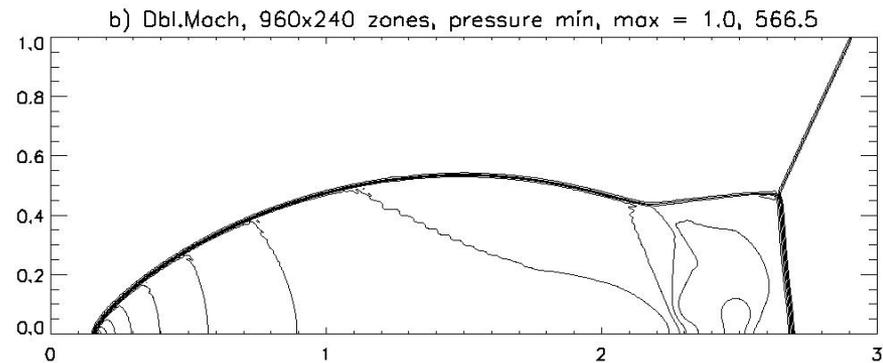
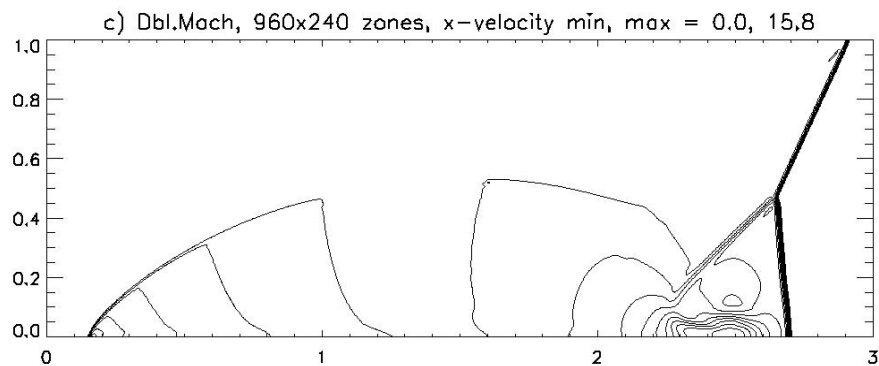
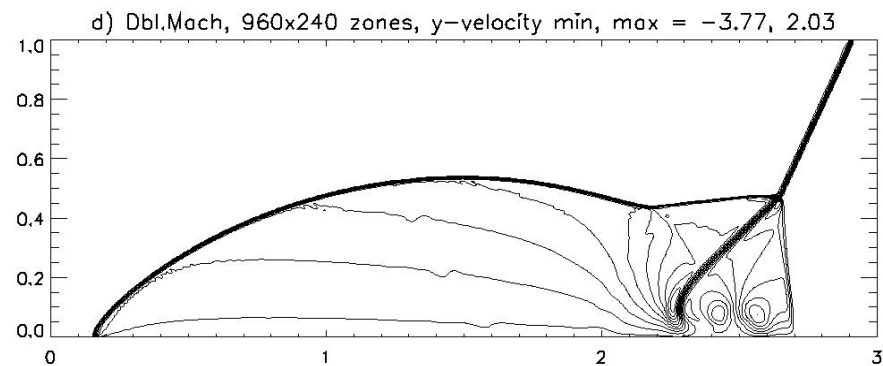

*Figs. 6a, 6b, 6c and 6d show contour plots of the final density, pressure, x-velocity and y-velocity respectively for the double Mach reflection problem. A resolution of 960×240 zones was used. We only show part of the mesh that spans [0,3]×[0,1]. 20 contours were used to show the density which ranges over [1.4,20.8]; the pressure which ranges over [1.0,566.5]; the x-velocity which ranges over [0.0,15.8] and the y-velocity which ranges over [-3.77,2.03].*

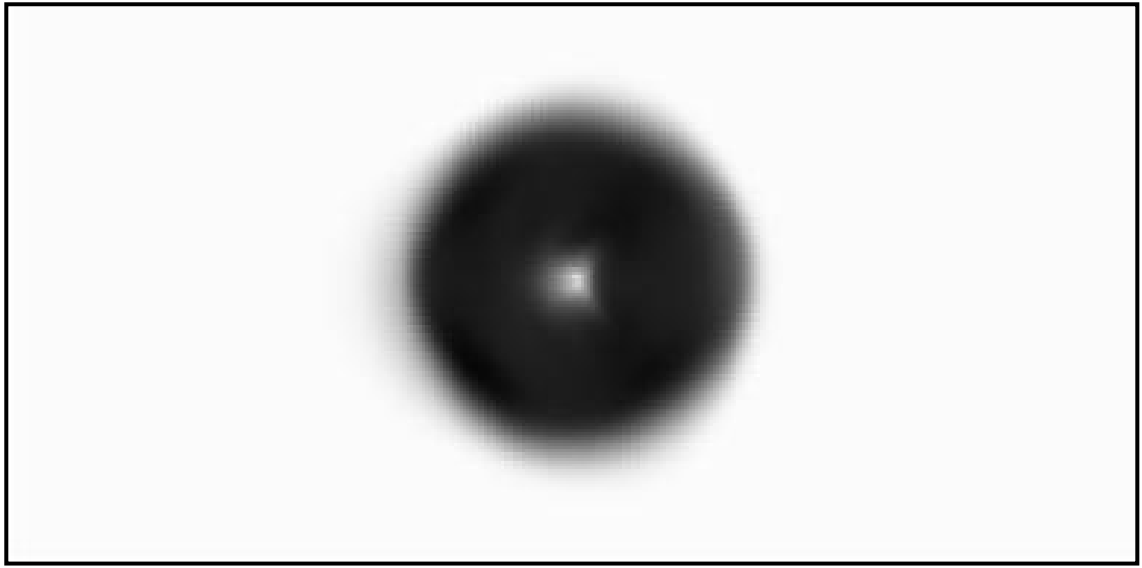

Fig. 7 shows the magnitude of the magnetic field as a grayscale image for the field loop advection problem. The loop is advected along the diagonal of the rectangular domain shown here. A 128×64 zone mesh was used. The plot shows the field loop after it has executed one complete orbit around the computational domain.

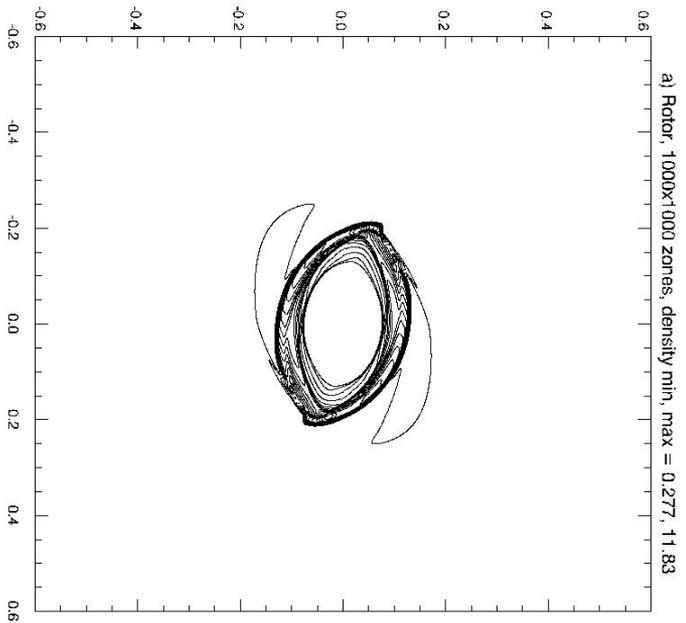

a) Rotor, 1000x1000 zones, density min, max = 0.277, 11.83

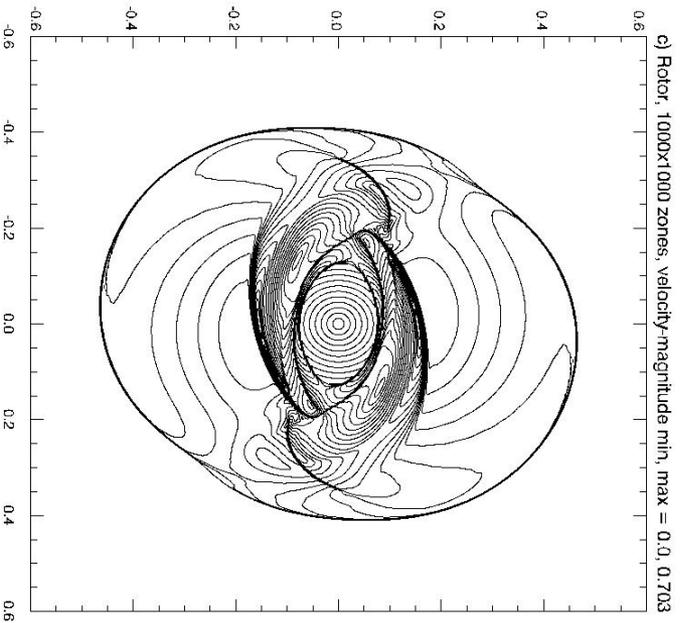

c) Rotor, 1000x1000 zones, velocity-magnitude min, max = 0.0, 0.703

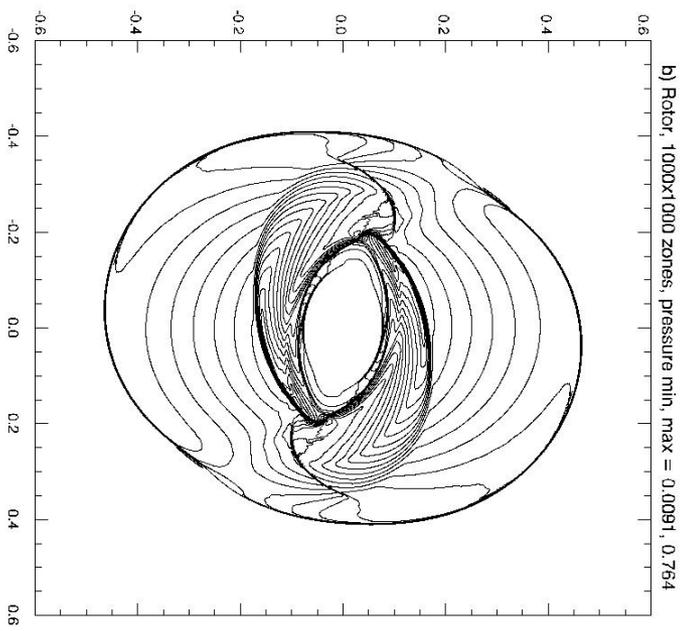

b) Rotor, 1000x1000 zones, pressure min, max = 0.0091, 0.764

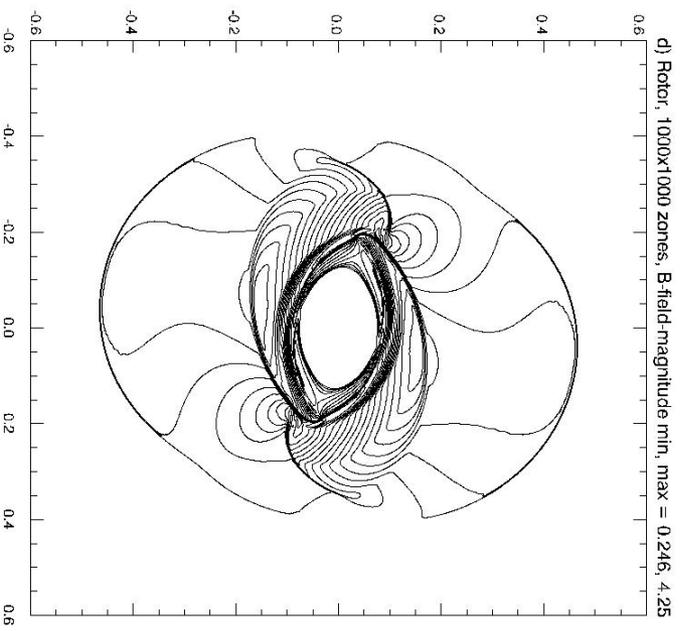

d) Rotor, 1000x1000 zones, B-field-magnitude min, max = 0.246, 4.25

*Figs. 8a, 8b, 8c and 8d show the final density, pressure, magnitude of the velocity and magnitude of the magnetic field for the rotor problem on a $1000^2$ zone mesh. 20 contours are shown in each plot. The density spans the range [0.277,11.83]; the pressure ranges over [0.0091,0.764]; the magnitude of the velocity has the range [0.0,0.703]; the magnitude of the magnetic field has the range [0.246,4.25].*

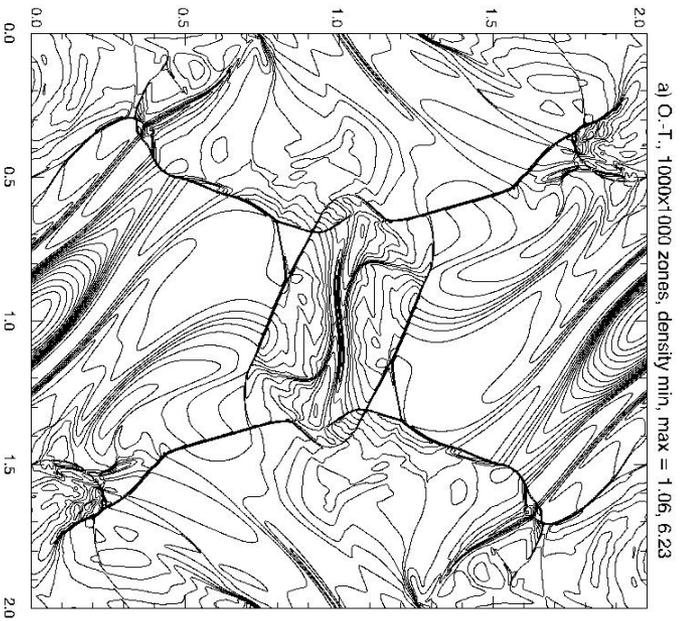

a) O.-T., 1000x1000 zones, density min, max = 1.06, 6.23

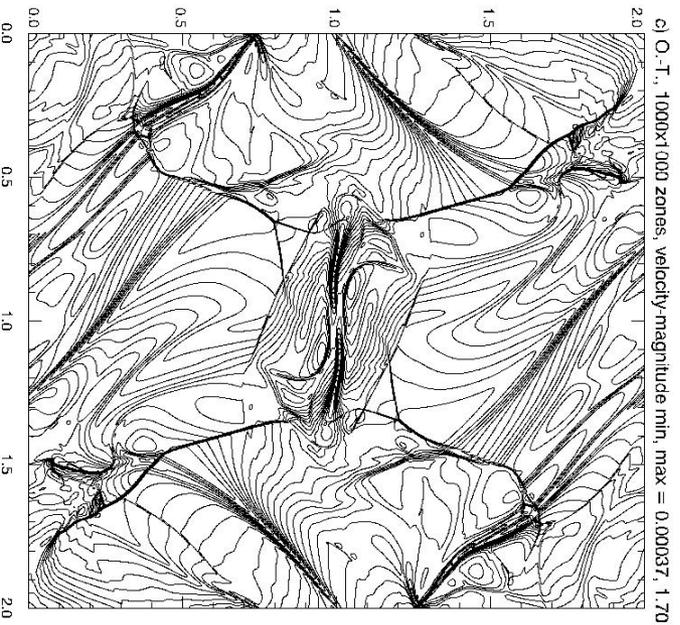

c) O.-T., 1000x1000 zones, velocity-magnitude min, max = 0.00037, 1.70

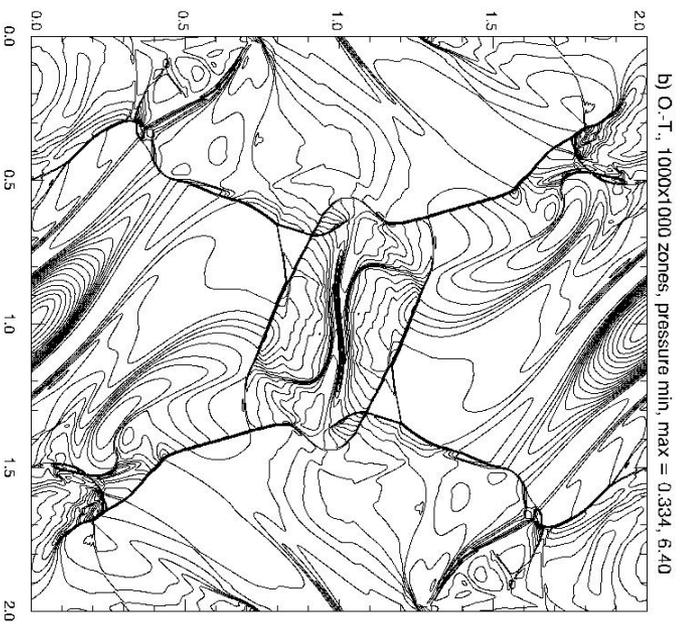

b) O.-T., 1000x1000 zones, pressure min, max = 0.334, 6.40

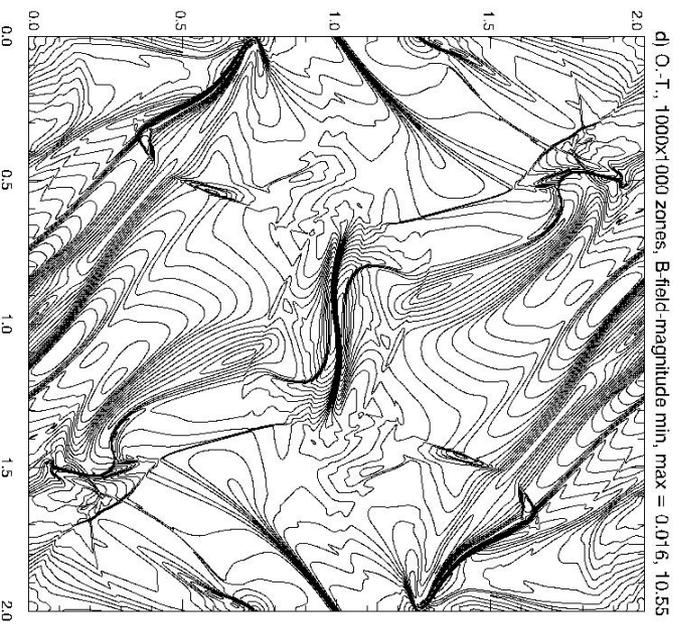

d) O.-T., 1000x1000 zones, B-field-magnitude min, max = 0.016, 10.55

*Figs. 9a, 9b, 9c and 9d show the final density, pressure, magnitude of the velocity and magnitude of the magnetic field for the Orzag-Tang problem on a $1000^2$ zone mesh. 20 contours are shown in each plot. The density spans the range [1.06,6.23]; the pressure ranges over [0.334,6.40]; the magnitude of the velocity has the range [0.00037,1.70]; the magnitude of the magnetic field has the range [0.016,10.55].*

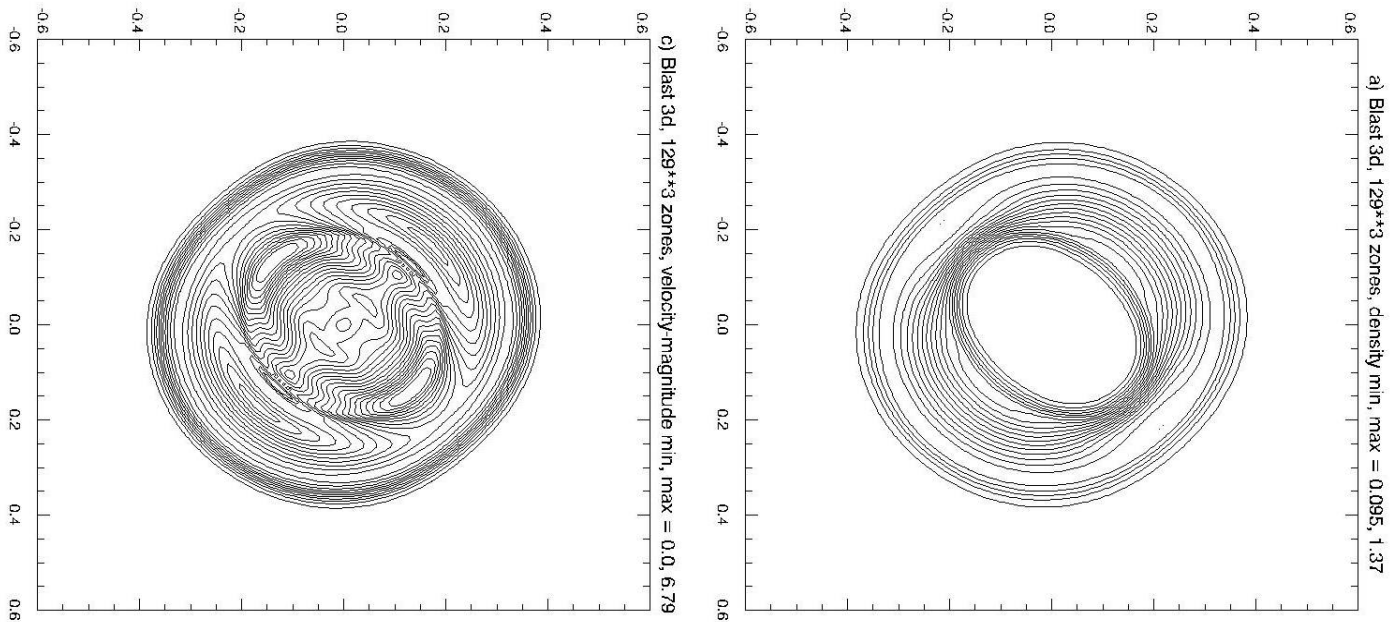

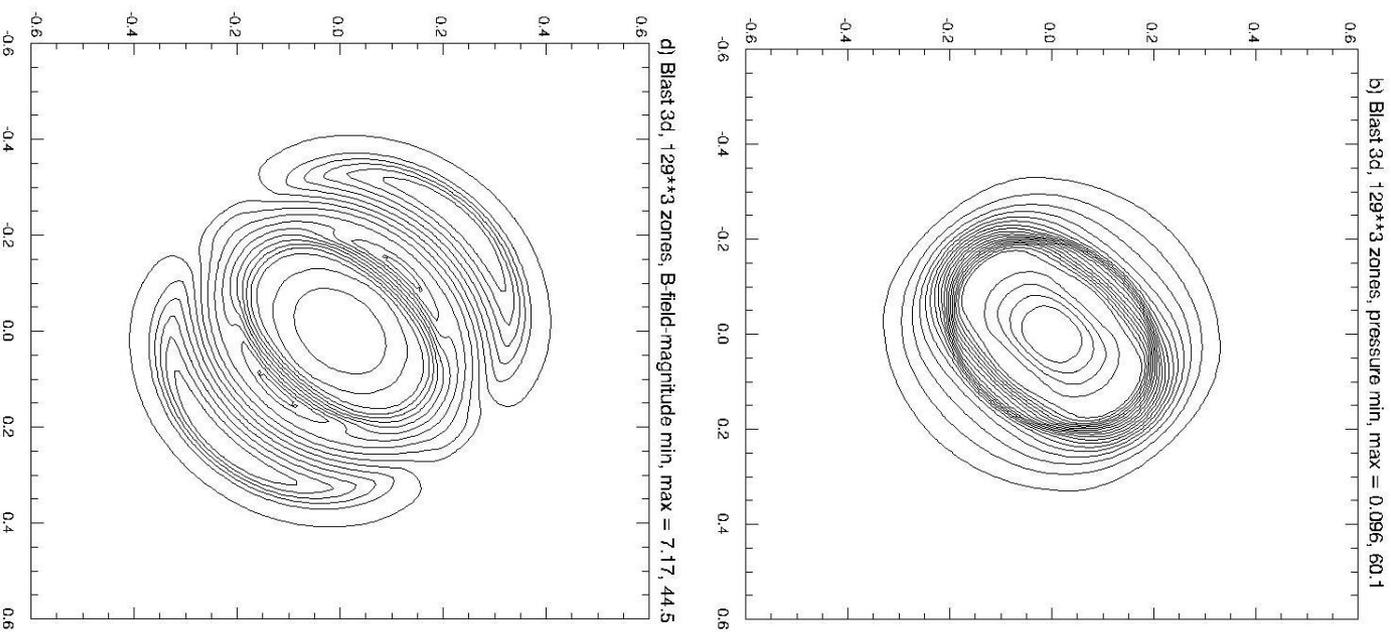

*Figs. 10a, 10b, 10c and 10d show the final density, pressure, magnitude of the velocity and magnitude of the magnetic field for the three dimensional magnetized blast problem on a $129^3$ zone mesh. The variables are shown in the xy-plane that passes through the middle of the domain. 20 contours are shown in each plot. The density spans the range [0.095,1.37]; the pressure ranges over [0.096,60.1]; the magnitude of the velocity has the range [0.0,6.79]; the magnitude of the magnetic field has the range [7.17,44.5].*